\documentclass{aa}
\usepackage{graphics}

\def\deg{\hbox{$^\circ$}}  
\def\arcmin{\hbox{$^\prime$}}  
\def\arcsec{\hbox{$^{\prime\prime}$}}

\def\fdg{\hbox{$.\!\!^\circ$}}  
\def\farcm{\hbox{$.\mkern-4mu^\prime$}}  
  
\def\NH{$N_{\rm HI}$} 
\def \hi {H\,{\sc i~}} 
 
\def\kms{km\,s$^{-1}$} 
 
\def\vlsr{$v_{\rm LSR}$}

\begin{document} 
\title{Arecibo imaging of compact high--velocity clouds} 
\titlerunning{Arecibo imaging of CHVCs} 
\author{W. B. Burton\inst{1}, R. Braun\inst{2}, and J.N. Chengalur\inst{3}} 
\institute{Sterrewacht Leiden, P.O. Box 9513, 2300 RA Leiden, The Netherlands 
\and Netherlands Foundation for Research in Astronomy, P.O. Box 2, 
7990 AA Dwingeloo, The Netherlands
\and National Centre for Radio Astrophysics, Post Bag 3, Ganeshkind PO, 
Pune, Maharashtra 411\,007, India } 

\date{Received mmddyy/ Accepted mmddyy} 
\offprints{W.B. Burton} 

\abstract{ Ten isolated compact high--velocity clouds (CHVCs) of the
  type cataloged by Braun \& Burton (\cite{brau99}) have been imaged
  with the Arecibo telescope and were found to have a nested core/halo
  morphology.  We argue that a combination of high--resolution
  filled--aperture and synthesis data is crucial to determining the
  intrinsic properties of the CHVCs. We identify the halos as Warm
  Neutral Medium surrounding one or more cores in the Cool Neutral
  Medium phase. These halos are clearly detected and resolved by the
  Arecibo filled--aperture imaging, which reaches a limiting
  sensitivity (1$\sigma$) of \NH\,$\sim 2 \times 10^{17}$ cm$^{-2}$
  over the typical 70~\kms~linewidth at zero intensity. The FWHM
  linewidth of the halo gas is found to be 25 \kms, consistent with a
  WNM thermal broadening within 10$^4$~K gas. Substantial asymmetries
  are found at high \NH\ ($>10^{18.5}$ cm$^{-2}$) levels in 60\% of our
  sample. A high degree of reflection-symmetry is found at low \NH\
  ($<10^{18.5}$ cm$^{-2}$) in all sources studied at these levels. The
  column--density profiles of the envelopes are described well by the
  sky--plane projection of a spherical exponential in atomic volume
  density, which allows estimating the characteristic central halo
  column density, \NH(0)~=~$4.1\pm3.2\times 10^{19}$ cm$^{-2}$, and
  characteristic exponential scale--length,
  $h_B~=~420\pm90$~arcsec. For plausible values of the thermal pressure
  at the CNM/WNM interface, these edge profiles allow distance
  estimates to be made for the individual CHVCs studied here which
  range between 150 and 850~kpc. An alternate method of distance
  estimation utilizing the mean exponential scale-length found in
  nearby low mass dwarf galaxies, $h_B~=~10.6\pm4.0$~kpc, yields
  distances in the range 320 to 730~kpc. A consequence of having
  exponential edge profiles is that the apparent size and total flux
  density of these CHVCs will be strongly dependent on the resolution
  as well as on the sensitivity of the data used; even a relatively
  deep observation with a limiting sensitivity of $\sim 10^{19}$
  cm$^{-2}$ over 70~\kms~will detect only the central 30\% of the
  source area and less than 50\% of the total flux density.  The
  exponential profiles also suggest that the outer envelopes of the
  CHVCs are not tidally truncated.  Several CHVC cores exhibit a
  kinematic gradient, consistent with rotation.  The halos appear
  kinematically decoupled from the cores, in the sense that the halos
  do not display the velocity gradients shown by the dense cores; the
  gradients are therefore not likely to be due to an external cause
  such as tidal shear. The much higher degree of symmetry observed in
  the halos relative to the cores also argues against an external cause
  of asymmetries in the cores.   }

\maketitle 

\section{Introduction} 
\label{sec:intro}

There are two principal categories of the anomalous \hi high--velocity
cloud phenomenon.  The first category, which contributes most of the
emission flux, consists of low--contrast maxima of extended diffuse
complexes with angular sizes up to tens of degrees. These complexes
contribute a large fraction of the total HVC flux density observed;
examples include the Magellanic Stream of debris from the Galaxy/LMC
interaction (e.g. Putman \& Gibson \cite{putm99}) as well as some
half--dozen extended HVC complexes (including complexes A, C, H, M, and
others). Although firm constraints on distances have been achieved only
for a few of the complexes, the available distance determinations show
these to lie rather nearby, at about 10 kpc (e.g. van Woerden et al.
\cite{woer99}). If, as seems plausible, the other large complexes also
lie at distances ranging from several to some 50 kpc, they will have
been substantially affected by the radiation and gravitational fields
of the Milky Way. The second category of anomalous \hi high--velocity
clouds are the compact high velocity clouds (CHVCs, Braun \& Burton
\cite{brau99}, hereafter BB99).  CHVCs are intrinsically compact,
isolated objects with angular sizes of about 1~degree. The spatial and
kinematic distributions of the CHVCs were found to be consistent with a
dynamically cold ensemble spread throughout the Local Group, but with a
net negative velocity with respect to the mean of the Local Group
galaxies. This net negative velocity would imply an infall towards the
Local Group barycenter at some 100 \kms.

The possibility that some of the high--velocity clouds might be
essentially extragalactic has been considered in various contexts by,
among others, Oort (\cite{oort66}, \cite{oort70}, \cite{oort81}),
Verschuur (\cite{vers75}), Eichler (\cite{eich76}), Einasto et al.
(\cite{eina76}), Giovanelli (\cite{giov81}), Bajaja et al.
(\cite{baja87}), Wakker \& van Woerden (\cite{wakk97}), BB99, and Blitz
et al. (\cite{blit99}).  Blitz et al. revived the suggestion that
high--velocity clouds are the primordial building blocks fueling
galactic growth and evolution.

It is plausible to hypothesize that the high--velocity clouds may be
viewed in terms of the hierarchical structure formation paradigm: the
large HVC complexes would be nearby objects currently undergoing
accretion onto the Galaxy, or representing tidal debris from a close
encounter, while the compact, isolated CHVCs would be their distant
counterparts, scattered throughout the Local Group environment.  The
accumulating support for such an hypothesis includes the evidence found
by Helmi et al. (\cite{helm99}) in the Milky Way halo for a recently
accreted dwarf galaxy, and the presence of the tidally disrupted
Sagittarius dwarf galaxy (Ibata et al.  \cite{ibat94}), which suggests
stellar analogies to the extended HVC complexes; the theoretical
simulations requiring numerous ``mini--halo" systems (Klypin et
al. \cite{klyp99}, Moore et al. \cite {moor99}); as well as the
relatively direct evidence regarding the distribution of some of the
anomalous--velocity gas (Blitz et al.  \cite{blit99}, BB99).  The
CHVCs, which have a spatial and kinematic deployment similar to that of
the Local Group dwarf galaxies, would represent inflowing material at
substantial distances. Indications of internal star formation have not
yet been found in CHVCs; if the objects are scattered throughout the Local
Group, they have probably not yet been exposed to the Galactic
radiation field or to strong external gravitational torques, and
consequently the material would be in a less evolved form.

\section{Motivation for filled--aperature observations}

BB99 identified and confirmed a sample of 65 CHVCs. These objects
evidently represent a more homogenous population than would a sample
which included any of the major HVC complexes.  The BB99 CHVC catalog
was based on survey data made with telescopes of modest resolution.
The primary source was the Leiden/Dwingeloo Survey (LDS) of Hartmann \&
Burton (\cite{hart97}), characterized by the angular resolution of
$36'$ provided by the Dwingeloo 25--meter telescope; at this resolution
the CHVCs are largely unresolved.

Of the sample of 65 CHVCs cataloged, eight have been subject to
high--resolution synthesis imaging.  Wakker \& Schwarz (\cite{wakk91b})
had used the Westerbork Synthesis Radio Telescope (WSRT) to show that
both CHVC\,114$-$06$-$466 and CHVC\,114$-$10$-$430 exhibit a core/halo
structure.  Subsequently Braun \& Burton (\cite{brau00}, hereafter
BB00) imaged six additional CHVC fields using the WSRT and showed that
these objects also have a characteristic morphology whereby one or more
quiescent, low--dispersion (linewidths in the range 2 to 10 \kms~FWHM)
compact cores (angular diameters typically 1 to 20 arcmin) are
distributed over a region of some tens of arcmin extent, and are
embedded in a diffuse, warmer halo of about one degree angular extent.
We note that Cram \& Giovanelli (\cite{cram76}) had earlier interpreted
data, taken towards parts of an extended high--velocity cloud using the
NRAO 300--foot and 140--foot telescopes, in terms of cold cores
enveloped in warmer gas. However, that interpretation was based on
the decomposition of complex line profiles into multiple Gaussian
components, rather than direct observation of the individual components.

The high angular resolution of the WSRT data constitutes an important
advantage in directly detecting the compact cores, but a particular
disadvantage is inherent in synthesis data in detecting the diffuse
halos. BB00 showed that the compact cores revealed so clearly in the
WSRT data accounted for as little as 1\% to only as much as 55\% of the
\hi line flux detected in the single--dish LDS observations.  The cores
typically cover only some 15\% of the source area. The high resolution
of the synthesis data also allowed unambiguous identification of the
core material with the cool condensed phase of the \hi--- the CNM ---
with kinetic temperatures near 100 K.  However the diffuse structures
extending over more than about 10 arcmin are not adequately imaged by
the interferometer because of the missing short--spacing information.

BB00 attempted to account for this missing
short--spacing information using the LDS material, but the sparseness of
the total--power data supported only a crude correction. However, the
data available for the CHVCs do suggest a characteristic two--phase
structure, with the cores of CNM shielded by a halo of warm diffuse
\hi--- the WNM --- with temperatures near 10$^4$ K (corresponding to a
thermal linewidth of about 24 \kms~FWHM).  Since the diffuse halos
could not be detected convincingly and directly by the synthesis
imaging, this nested geometry remained conjecture for the targeted
objects. Direct detection of the diffuse halos was the principal
motivation for undertaking the Arecibo observations reported here:
these observations now provide the first resolved detection of the
diffuse halos of the CHVCs, and confirm that the halos are the WNM
gaseous phase providing a shielding column density for the CNM of the
compact cores.

The Arecibo telescope is well--suited to provide sensitivity to the
total column density at relatively high angular resolution.  This issue
is particularly important for CHVC targets, because they are of a size
comparable to the primary beam of most synthesis instruments. The
Arecibo data thus complement the interferometric data in a crucial
manner.

\setlength{\tabcolsep}{4pt}
\begin{flushleft} 
\begin{table*} 
\caption[]{Compact, isolated high--velocity clouds observed at Arecibo.
Column 1 gives the object designation; columns 2 and 3 give the
celestial coordinates (J2000); column 4 gives the {\sc rms} sensitivity
measured over 10 \kms~on the wings of the constant--declination
cross--cut; column 5 gives the velocity range over which the total
\NH~was determined in this cross--cut; and column 6 gives the
\NH~sensitivity, corresponding to the indicated {\sc rms} brightness
and the velocity integration range. }
\label{tab:sample}
\begin{tabular}{cccrcc} 
\noalign{\smallskip} \hline 
Name & RA(2000) & Dec(2000) & {\sc rms}\,(10\,\kms) & $\Delta v$ (\NH\,cut) &
\NH\,{\sc rms} 
\\ CHVC\,$lll \pm bb \pm vvv$ &(h m) & $(\degr~\arcmin~\arcsec)$ 
& (mK)~~~~~ & (\kms) & ($10^{17}\,{\rm cm}^{-2}$) \\
\noalign{\smallskip} \hline 
CHVC\,092$-$39$-$367 & 23 14.0 & 17 24 00 & 6.5~~~~~~ & 86 & 3.4 \\
CHVC\,100$-$49$-$383 & 23 50.5 & 11 19 00 & 4.6~~~~~~ & 63 & 2.1 \\
CHVC\,148$-$32$-$144 & 02 26.5 & 26 22 30 & 11.0~~~~~~ & 63 & 5.0 \\
CHVC\,158$-$39$-$285 & 02 41.5 & 16 17 30 & 3.8~~~~~~ & 86 & 2.0 \\
CHVC\,186$-$31$-$206 & 04 14.0 & 06 36 19 & 14.0~~~~~~ & 58 & 6.1 \\
CHVC\,186+19$-$114   & 07 17.5 & 31 52 00 & 5.9~~~~~~ & 75 & 2.9 \\
CHVC\,198$-$12$-$103 & 05 42.0 & 07 54 00 & 3.9~~~~~~ & 75 & 1.9 \\
CHVC\,202+30+057     & 08 27.0 & 21 46 00 & 7.0~~~~~~ & 65 & 3.2 \\
CHVC\,204+30+075     & 08 27.0 & 20 01 30 & 5.4~~~~~~ & 72 & 2.6 \\
CHVC\,230+61+165     & 10 55.0 & 15 28 30 & 5.8~~~~~~ & 80 & 3.0 \\
\noalign{\smallskip} \hline 
\end{tabular} 
\end{table*} 
\end{flushleft}

\section{Sample selection}
\label{sec:sam}

Ten examples of the compact, isolated high--velocity clouds of the type
cataloged by BB99 were selected for the Arecibo observations.  The
targets are listed by their CHVC designation in Table \ref{tab:sample}.
Nine of the sources had been identified in the BB99
compilation.  The tenth source, CHVC\,186$-$31$-$206, is evident
in the LDS but it had not been included in the BB99 list
because it appears in the same general area of the sky as the
Anticenter Stream complex and therefore had been excluded by the
isolation criterion of BB99.  The new, higher--quality Arecibo data suggest
that this source is sufficiently compact and isolated in velocity to be
placed in the CHVC category.  Three of the ten objects selected also
occur in the Wakker \& van Woerden (\cite{wakk91c}) catalog, namely
CHVC\,158$-$39$-$285 (WvW486), CHVC\,186+19$-$114
(WvW\,215), and CHVC\,198$-$12$-$103 (WvW\,343), the others
having evidently passed undetected through the relatively coarse
gridding lattice of the data on which the Wakker \& van Woerden catalog
was based.  The targets selected span both negative and positive radial
velocities, and occur in both the northern and southern Galactic
hemispheres.  Consistent with the operational definition of the CHVC
class of objects, the sources observed at Arecibo were only
marginally resolved in the 36--arcmin beam of the Dwingeloo telescope.

\section{Observations}
\label{sec:obs}
\subsection{Instrumental parameters}

The observations were carried out during seven days in November, 1999,
using the Gregorian feed together with the narrow L--band  (LBN) receiver.

The spherical primary of the Arecibo telescope is 305 meters in
diameter; the Gregorian optics comprises two subreflectors,
illuminating the primary over an area of about $210 \times 240$ meters
in extent.  The FWHM beamwidth measured with the wide L--band (LBW) feed at
1420 MHz is $3.1 \times 3.7$ arcmin in the azimuth and zenith--angle
directions, respectively (Heiles \cite{heil99}).  Measurements for the LBN
feed (Howell \cite{howe00}) yield values consistent with that of the
LBW feed.  Different sections of the spherical reflector are
illuminated, however, depending on the source position. In order to
minimize beam distortions and gain variations caused when the
illuminated pattern spills over the edges of the primary surface, we
constrained most of the observations to moderate zenith angles, less
than $17\deg$.  The pointing accuracy of the telescope system is about
$5''$, and thus of no concern for the extended sources observed in our
program.  The observations were carried out during the period extending
from local sunset until about two hours after sunrise, shown by
experience to provide the most stable baselines.

The spectrometer was a 9--bit 2048 channel autocorrelator observing two
polarizations with two simultaneous bandpass settings, namely 6.25 MHz
and 1.56 MHz, yielding $\Delta v = 1.3$ and 0.32 \kms, respectively.
The bands were centered on the \vlsr~of the CHVC targets as determined
by BB99 for 9 of the 10 targets, and from observations from the LDS for
CHVC\,186$-$31$-$206.

\subsection{Observing and calibration strategy} 

Each of the CHVC targets was first mapped on a grid of $1\deg \times
1\deg$ size on a fully--sampled 90 arcsec lattice, but using short
integrations, in order to determine the locations of the peak flux
concentrations and derive a nominal gain calibration for the field
based on the background continuum sources.  Each Nyquist--sampled
$1\deg \times 1\deg$ map involved 73 minutes of effective integration
time.  The average of the first and last 90--arcsec of RA of each
data-scan were used to calibrate the passband shape of each spectrum in
that scan. In the event that source emission extends to the edges of
the sampled region in RA, this will lead to a weighting down of such
features. This possibility must be borne in mind in the subsequent
analysis. An estimate of the continuum emission was then constructed by
averaging in frequency over the line-free data. Spatial Gaussian fits
were made to compact and unconfused continuuum sources which were
present by chance in the observed fields. Comparison with the same
sources detected in the NVSS (Condon et al. \cite{cond98}) allowed
determination of a nominal absolute gain calibration factor for each
field.  Typically, several suitable sources with flux densities in the
range of 100--200 mJy were present. The derived noise--equivalent flux
density (NEFD, or $T_{\rm Sys}$/Gain) averaged over all observed fields was
3.30$\pm$0.26~Jy. While a small systematic variation (of about 5\%) in
NEFD is expected with zenith angle, we chose to adopt the NEFD derived
from each fully--sampled image to calibrate all of the subsequent
drift--scan data aquired for that field.  The relationship between flux
density and brightness temperature follows from the beam area at
1420~MHz, $S$(Jy/Beam)~=~0.0634~$T_{\rm B}$(K).

These shallow images then served as finding charts on which to identify
the principal flux concentrations.  Longer--integration spectra were
then accumulated in a single cross--cut made at constant declination by
repeating drift scans of $2\deg$ length centered on this peak.  As many
as 75 constant--declination driftscans were accumulated for some of the
objects, providing integration times of as much as 15 minutes per beam.
As with the mapping data, the first and last 90~arcsec of each scan in
RA were averaged to perform the initial calibration of the passband
shape. A spatial smoothing in the RA direction with a 180~arcsec
Gaussian was employed to enhance the signal--to--noise figure with only a
modest degradation of spatial resolution. Various smoothings were employed in
the velocity direction to enhance detection of low--surface--brightness
emission features. On the basis of the final averaged and smoothed
data, the off--source ranges of RA and velocity were determined. The
off--source range of RA was used to form an average spectrum for a final
passband calibration. The off--source range of velocity was used to
determine the average continuum level to subtract from each spectrum.
No other baseline manipulation was employed.

The resulting {\sc rms} sensitivities over 10 \kms, as indicated in
column 4 of Table \ref{tab:sample}, were as low as 4 mK; this limit 
corresponds to sensitivity to
\hi column density, after summing over the typical, 70 \kms, emission
linewidth at zero intensity, of $2
\times 10^{17}$ cm$^{-2}$.  The resulting \hi material constitutes the
most sensitive yet obtained for high--velocity clouds.  This
sensitivity is particularly important for determining the properties of
the diffuse halos, largely inaccessible in synthesis data.

\subsection{Observational displays}

The observational material is displayed as follows.

\begin{figure*}
\resizebox{16cm}{!}{\includegraphics{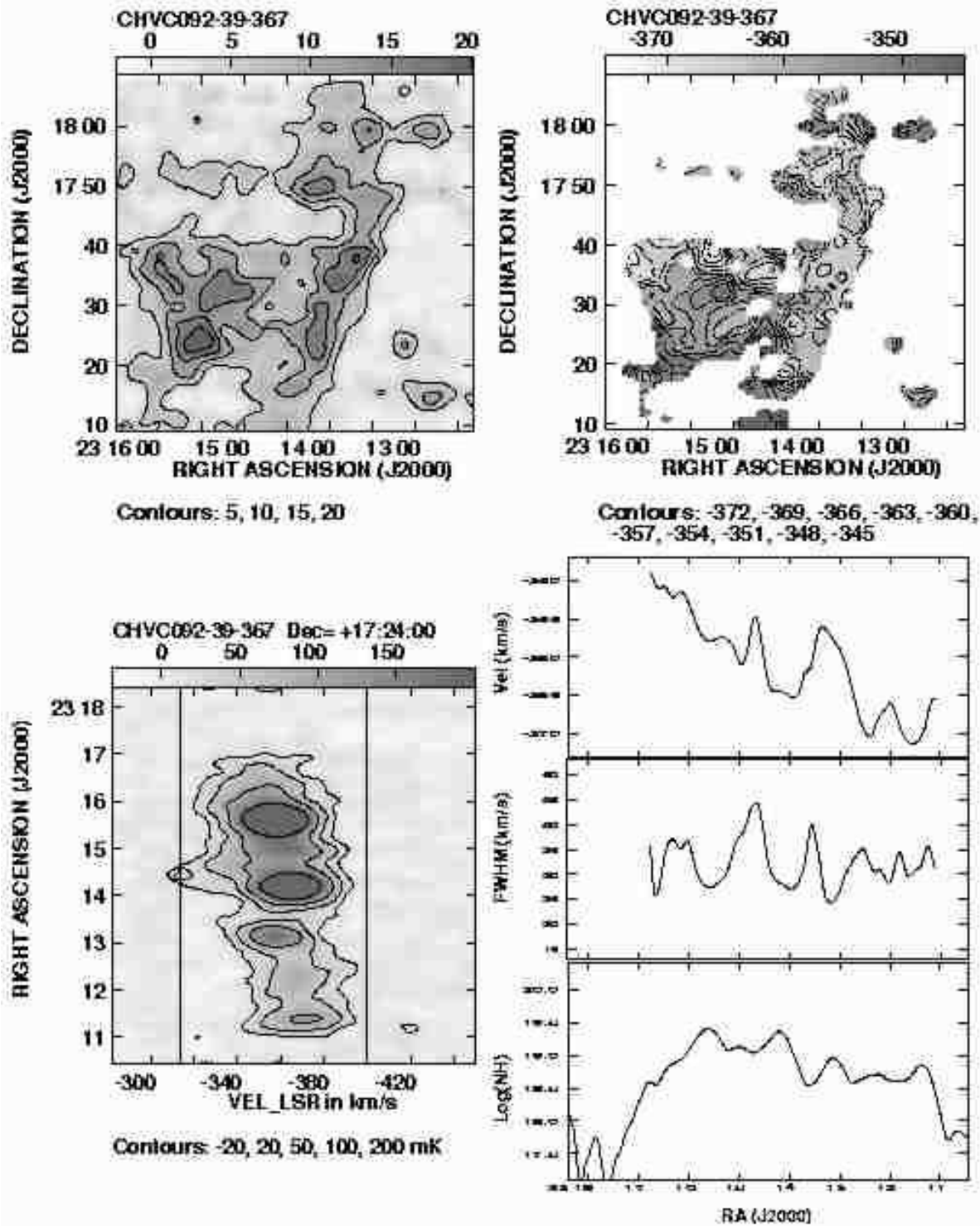}}
\caption{ Imaging and cross--cut data observed for CHVC\,092$-$39$-$36.
Upper left: \NH~distribution over the $1^\circ \times 1^\circ$ grid
Nyquist sampled in short integrations; the indicated contours represent
units of $10^{18}$\,cm$^{-2}$.  The right ascension labeling refers to
the tick mark above the last zero of the label. Upper right:
Intensity--weighted velocity field across the mapped grid; the contours
represent \vlsr in units of \kms.  Lower left: $\alpha$,\,\vlsr~slice
sampled in longer integrations at the indicated declination; the
contours represent $T_{\rm B}$ in units of mK.  Lower right: Centroid
\vlsr, velocity FWHM, and \NH~determined along the
$\alpha$,\,\vlsr~slice, within the velocity limits indicated by the
vertical lines in the panel on the lower left.  }
\label{fig:h092}
\end{figure*}

\begin{figure*}
\resizebox{16cm}{!}{\includegraphics{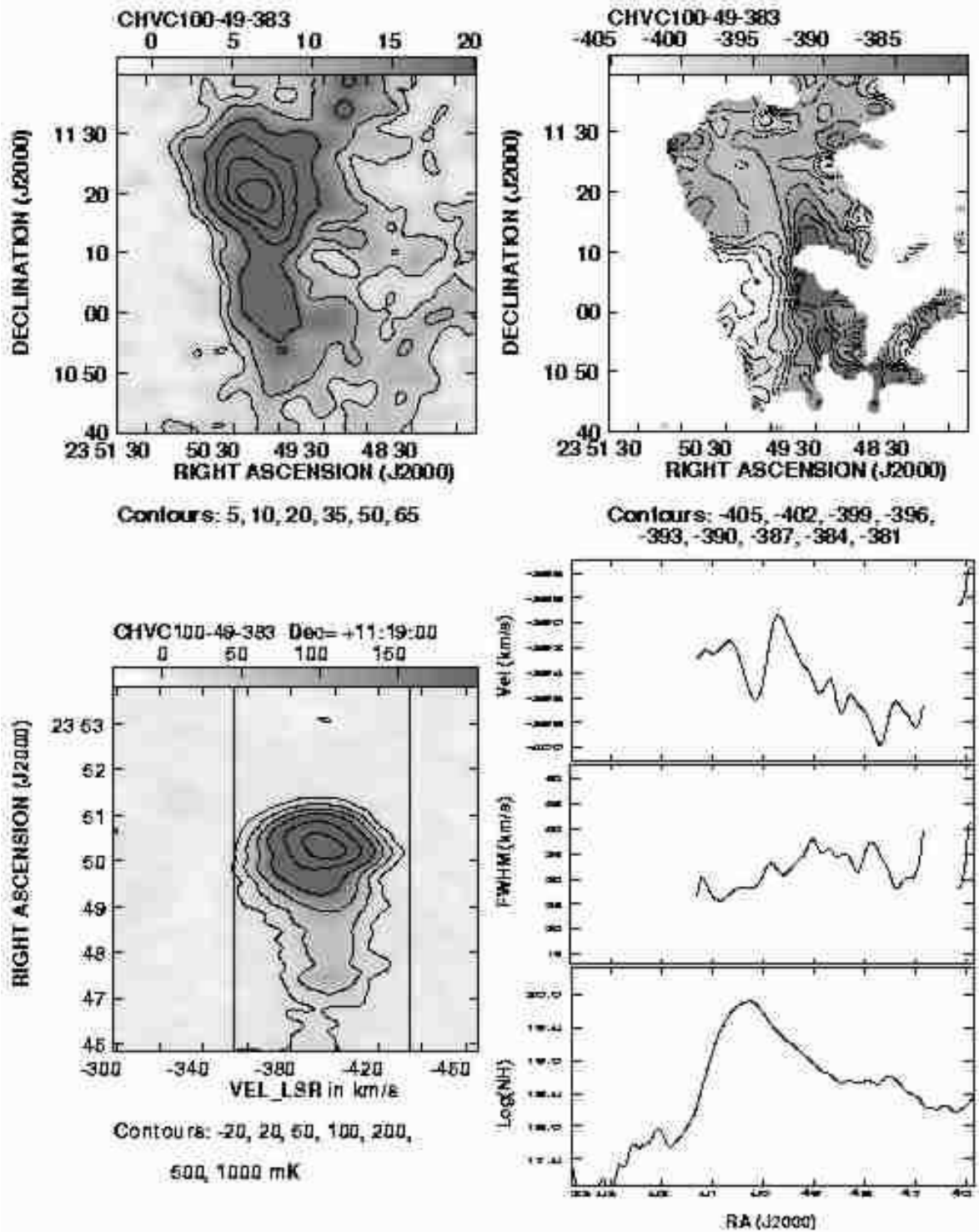}}
\caption{ 
   Imaging and cross--cut data observed for CHVC\,100$-$49$-$383
   as in Fig.~\ref{fig:h092}.}
\label{fig:h100}
\end{figure*}

\begin{figure*}
\resizebox{16cm}{!}{\includegraphics{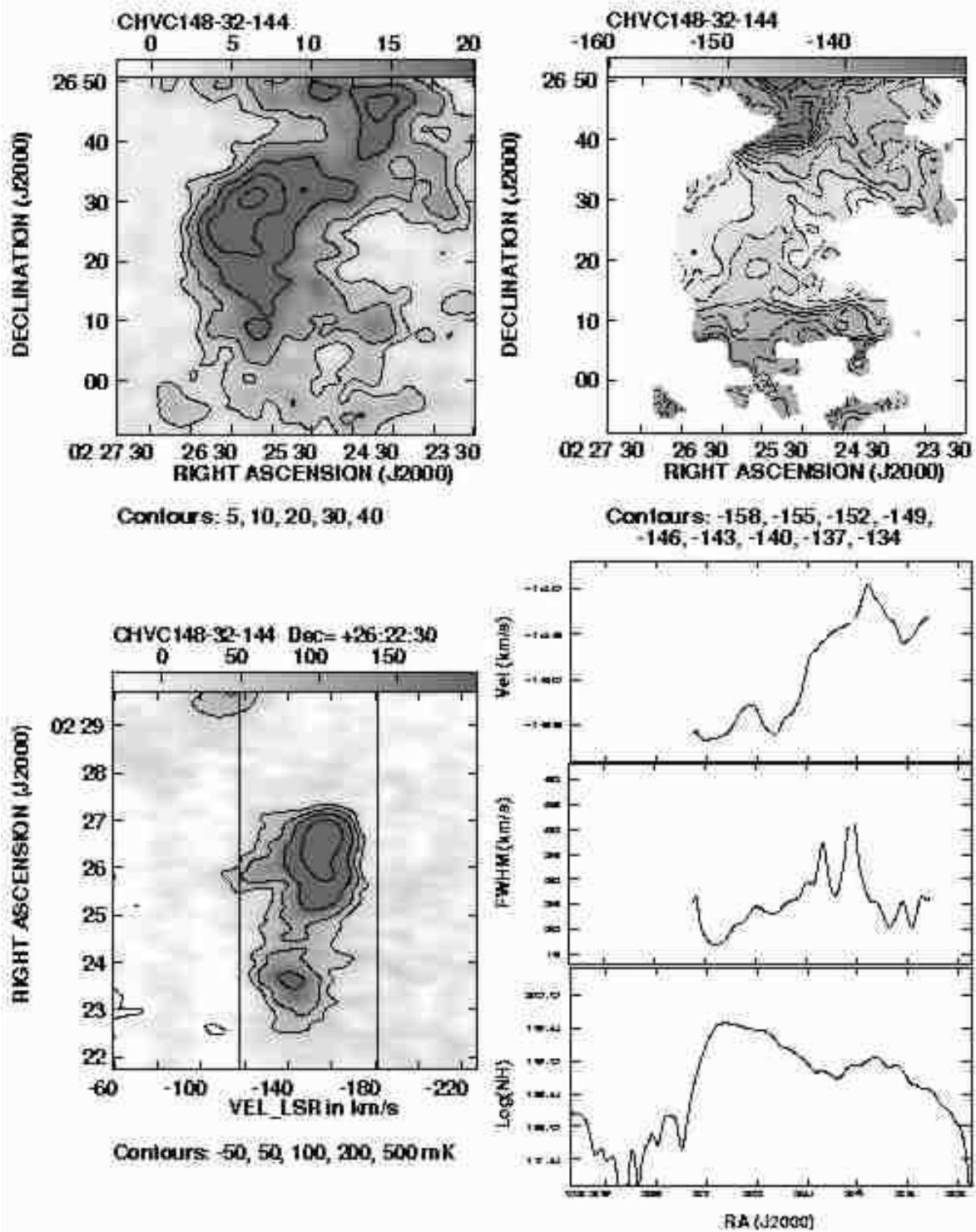}}
\caption{
  Imaging and cross--cut data observed for CHVC\,148$-$32$-$144 
   as in Fig.~\ref{fig:h092}.}
\label{fig:h148}
\end{figure*}

\begin{figure*}
\resizebox{16cm}{!}{\includegraphics{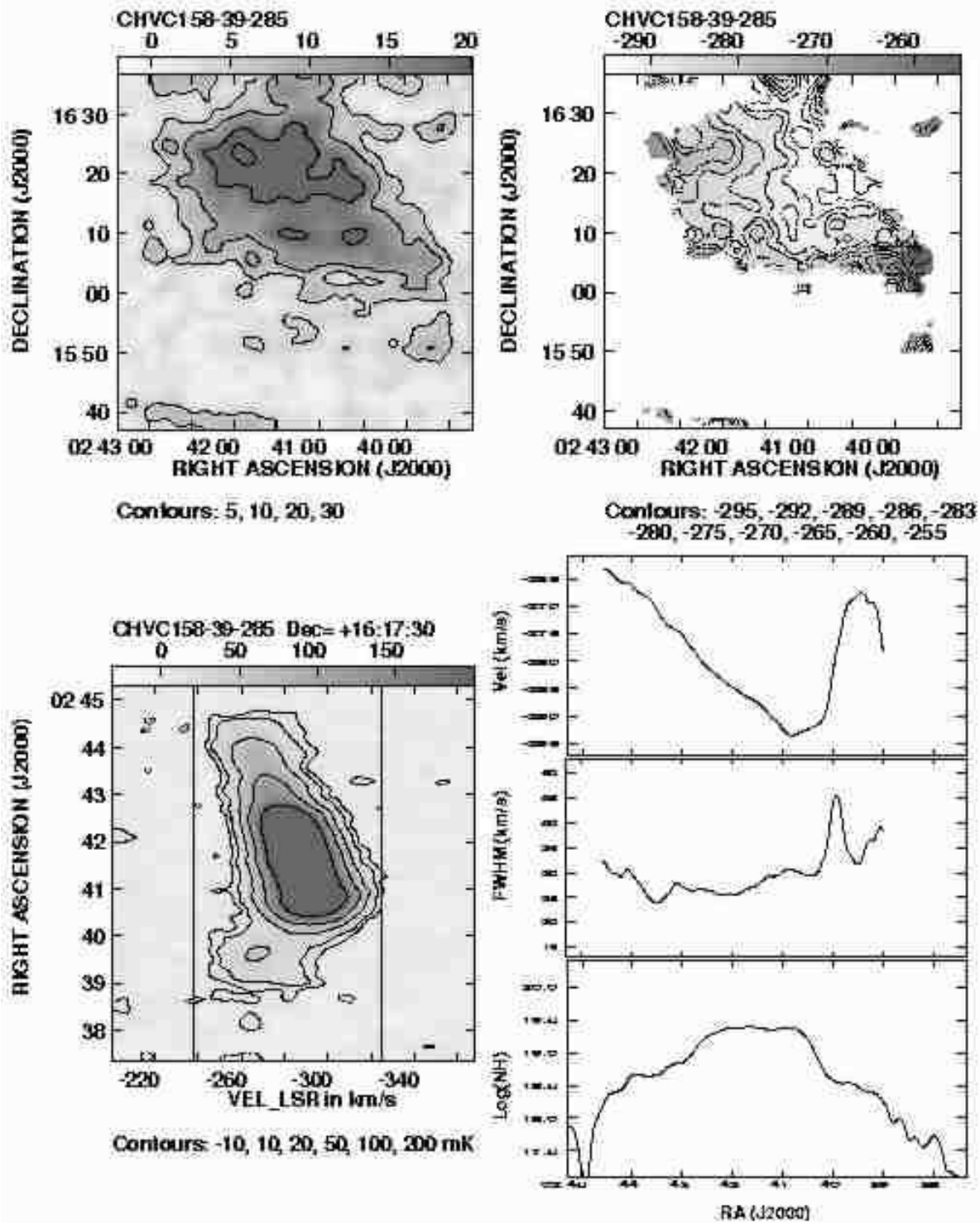}}
\caption{
  Imaging and cross--cut data observed for CHVC\,158$-$39$-$285
   as in Fig.~\ref{fig:h092}.}
\label{fig:h158}
\end{figure*}

\begin{figure*}
\resizebox{16cm}{!}{\includegraphics{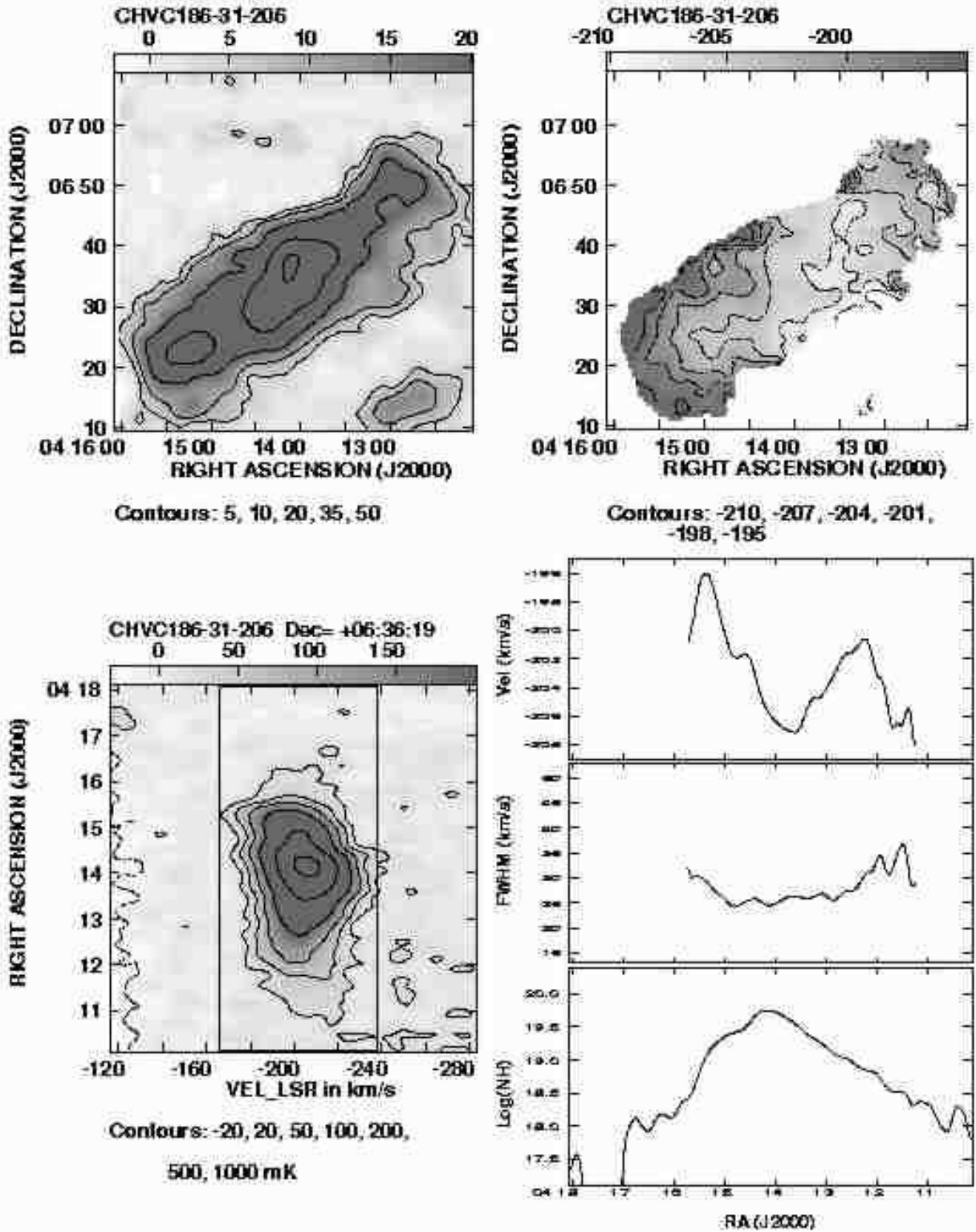}}
\caption{
  Imaging and cross--cut data observed for CHVC\,186$-$31$-$206 
   as in Fig.~\ref{fig:h092}.}
\label{fig:a186}
\end{figure*}

\begin{figure*}
\resizebox{16cm}{!}{\includegraphics{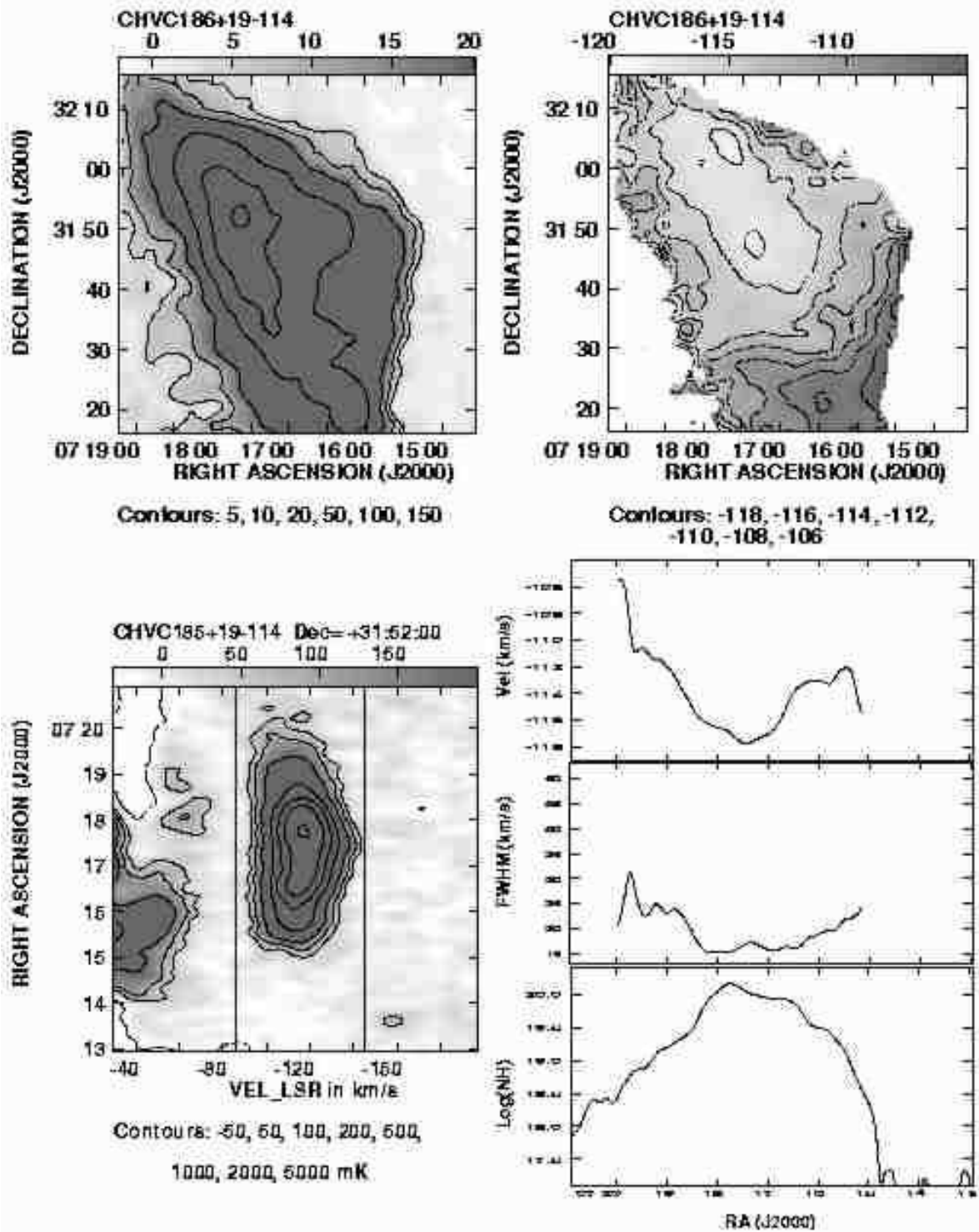}}
\caption{
  Imaging and cross--cut data observed for CHVC\,186+19$-$114
   as in Fig.~\ref{fig:h092}.}
\label{fig:h186}
\end{figure*}

\begin{figure*}
\resizebox{16cm}{!}{\includegraphics{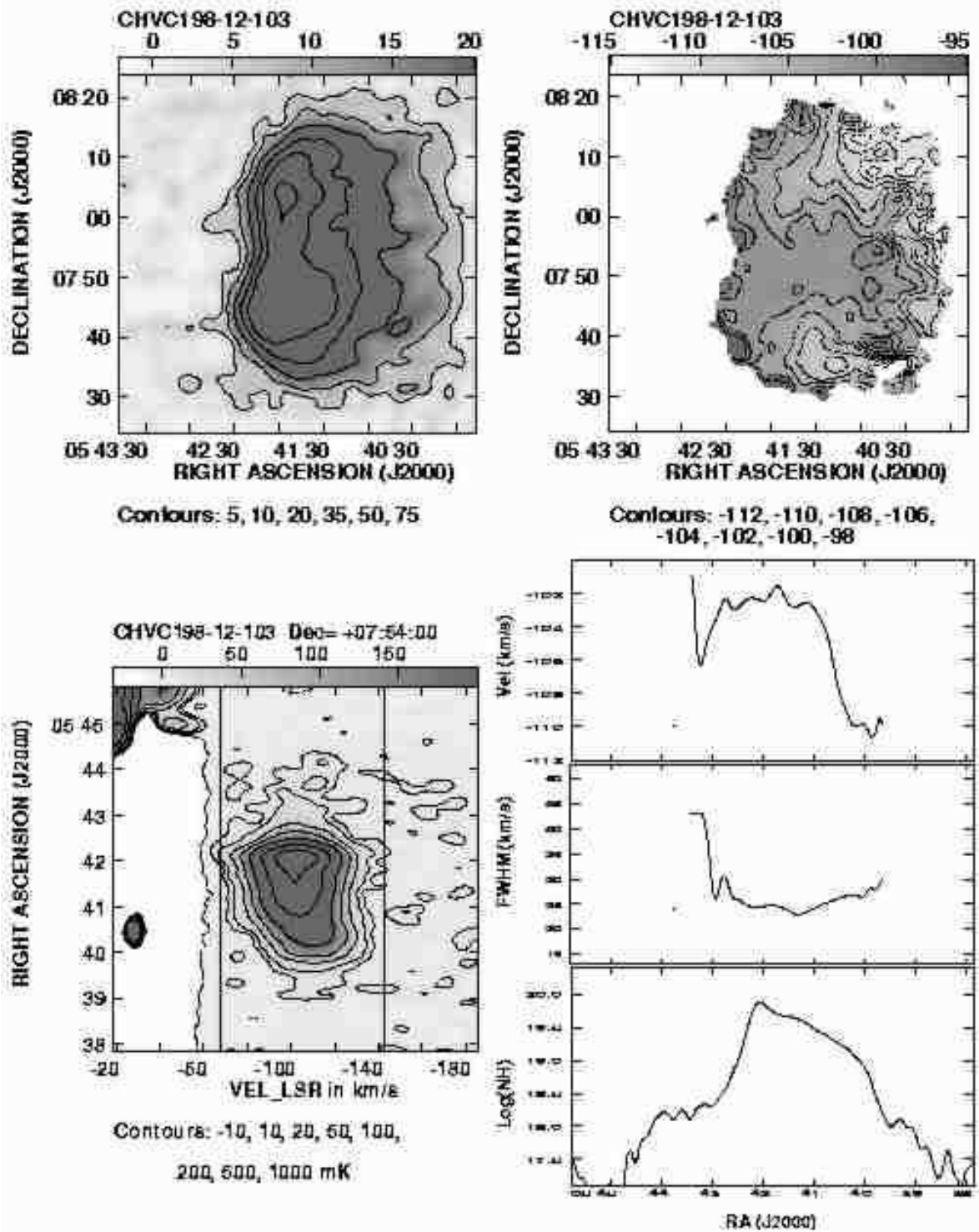}}
\caption{
  Imaging and cross--cut data observed for CHVC\,198$-$12$-$103 
   as in Fig.~\ref{fig:h092}.}
\label{fig:h198}
\end{figure*}

\begin{figure*}
\resizebox{16cm}{!}{\includegraphics{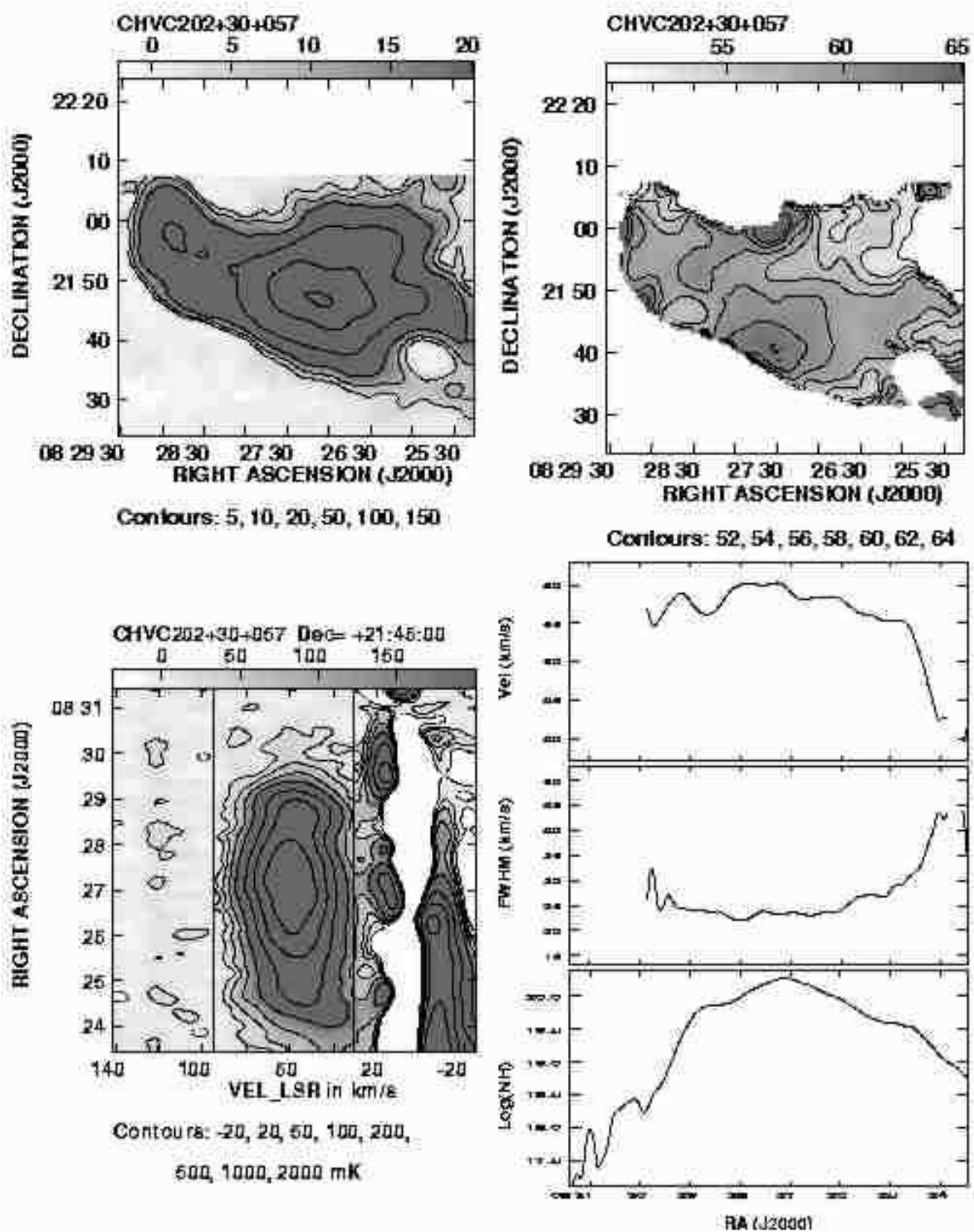}}
\caption{
  Imaging and cross--cut data observed for CHVC\,202+30+057
   as in Fig.~\ref{fig:h092}.}
\label{fig:h202}
\end{figure*}

\begin{figure*}
\resizebox{14.5cm}{!}{\includegraphics{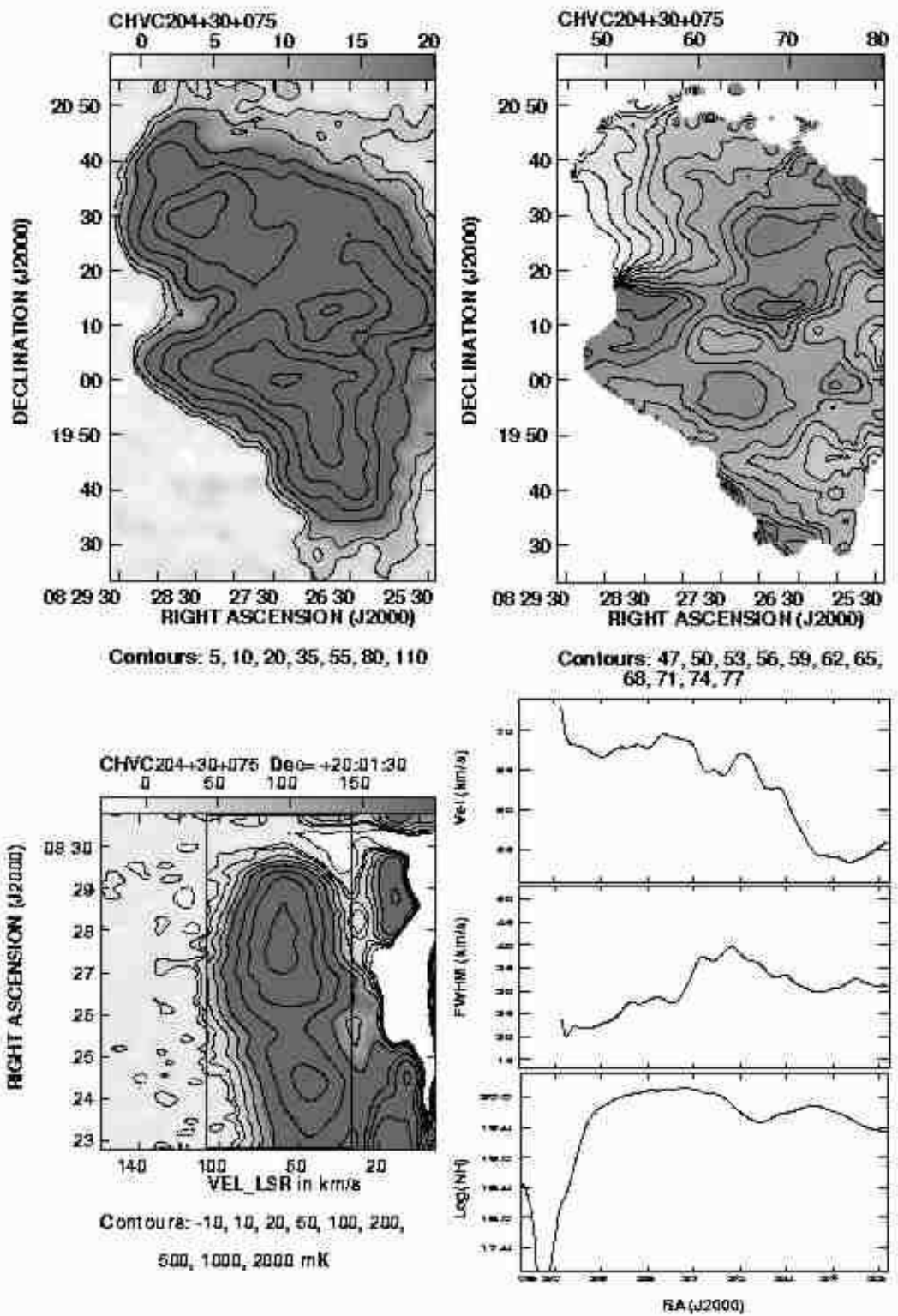}}
\caption{
  Imaging and cross--cut data observed for CHVC\,204+30+075
  over the $1^\circ \times 1.^\circ5$ grid as in Fig.~\ref{fig:h092}.}
\label{fig:h204}
\end{figure*}

\begin{figure*}
\resizebox{16cm}{!}{\includegraphics{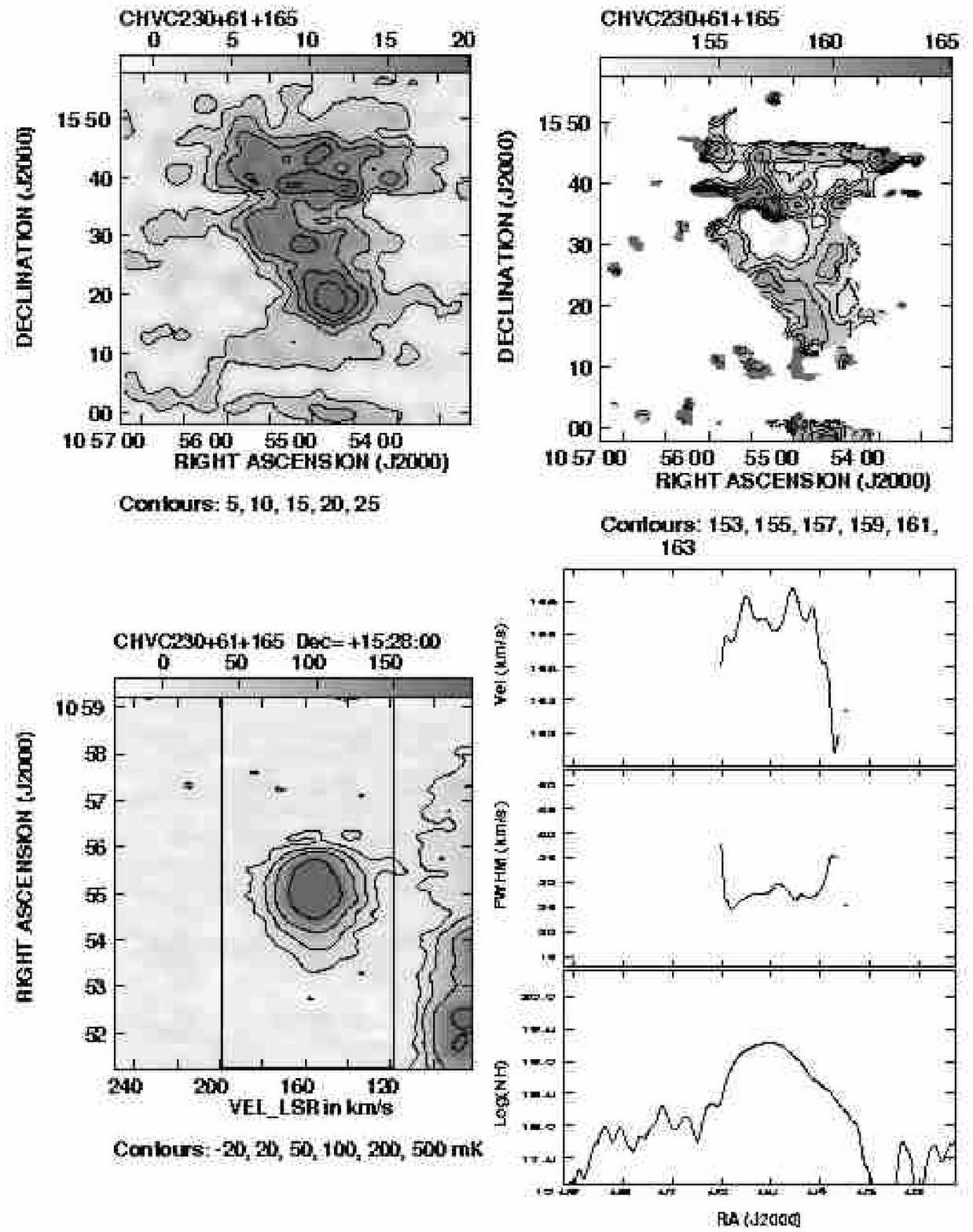}}
\caption{
  Imaging and cross--cut data observed for CHVC\,230+61+165
   as in Fig.~\ref{fig:h092}.}
\label{fig:h230}
\end{figure*}

The panel on the upper left in each of Figs. \ref{fig:h092} through
\ref{fig:h230} shows contours of integrated intensity for the shallow
$1\deg \times 1\deg$ images made of each of the CHVC targets.  These
moment--map panels show the integrated \hi column depth at the contour
levels, in units of $10^{18}$ cm$^{-2}$, indicated below each panel,
with the range of integration given in the column 5 of Table
\ref{tab:sample}.  The contour plot on the upper right in each of Figs.
\ref{fig:h092}--\ref{fig:h230} shows the intensity--weighted
velocity field as determined from the data in the integrated--flux map.
The contours give the intensity--weighted \vlsr, at the levels
indicated below the panel.

The panels in the lower portion of the various Figs.~\ref{fig:h092}
through \ref{fig:h230} refer to the deep driftscan material accumlated
over $2\deg$ at a central declination chosen for each CHVC.  The panel
on the lower left in each of the figures shows the resulting
position,\,velocity map for each target at a velocity smoothing of 10
\kms~FWHM.  The constant declination along which the deep driftscan was
made is indicated above each of these panels; the contours give the
intensities in units of mK, at the levels indicated below each panel.
The panels adjacent to the $\alpha$,\,\vlsr~cuts show (from upper to
lower, respectively) the \vlsr~of the emission centroid measured along
the cut, the velocity FWHM of the emission, and the logarithm of
\NH~measured along the constant--$\delta$ cross--cut.  These properties
were determined from the region indicated by the vertical lines in the
adjacent position,\,velocity diagram.  This velocity range was chosen
to encompasses as much as possible of the detected \hi emission from
each object while excluding any confusing features.

Several of the position,\,velocity cross--cuts display a kinematic
gradient, but it is important to remain aware that the
longer--integration cross--cuts refer to a single slice, in a specific
orientation, through a particular emission concentration, across a CHVC
which may in fact comprise multiple cores.  Until the data can be improved
such that the entire object is imaged deeply, the shallow images shown in the
upper panels of the relevant figure must be consulted to judge if this
gradient is aligned with a possible elongation of the spatial map or with
possible kinematic gradients seen in the larger context.

\begin{figure*}
\resizebox{14cm}{!}{\includegraphics{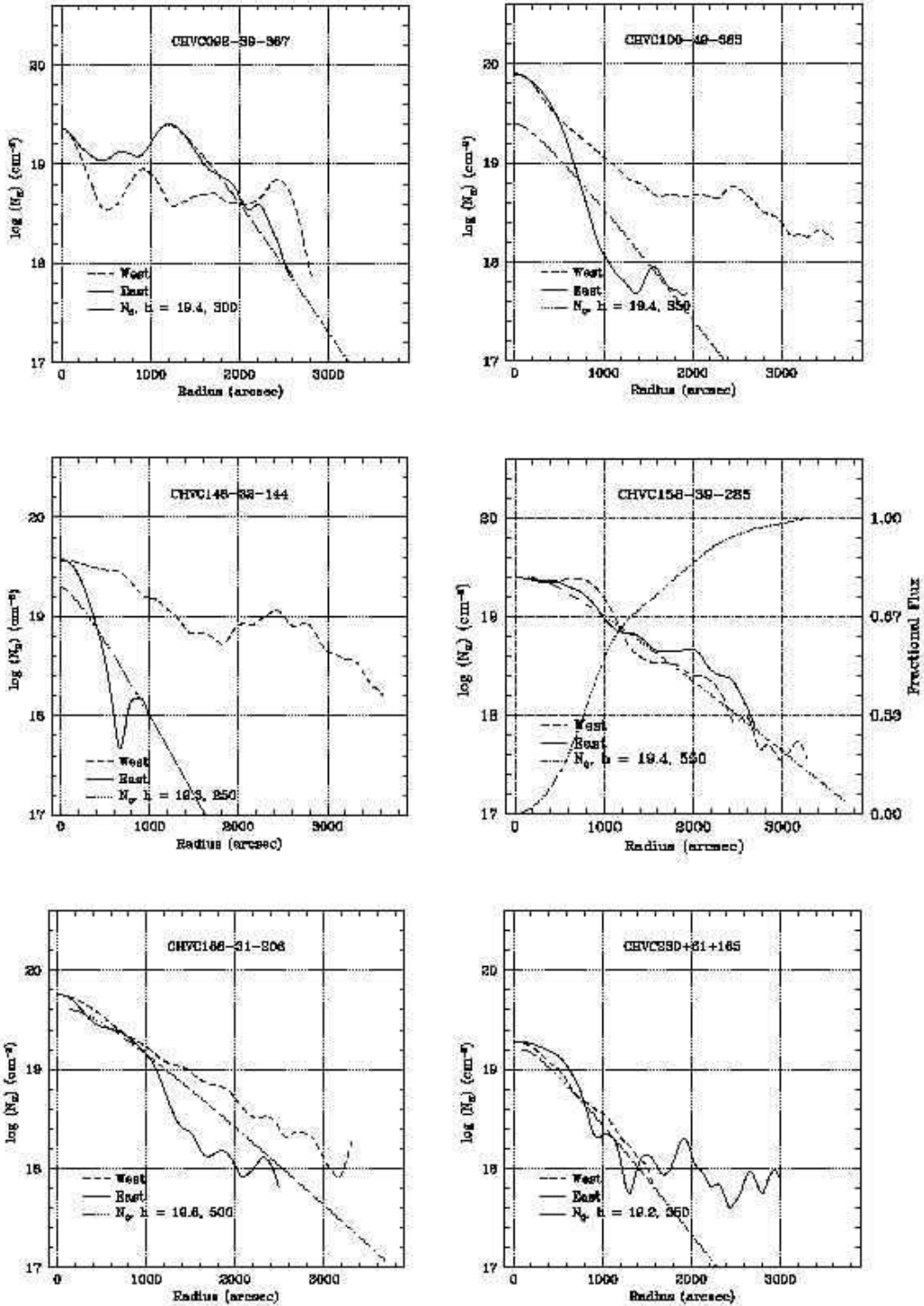}}
\caption{
  Column--density profiles of the indicated CHVCs. The logarithm
  of \hi column density is plotted against distance to the East and
  West of the emission peak in the lower--left panel of
  Figs.~\ref{fig:h092}--\ref{fig:h230}. The dotted curve overlaid on
  the observed log(\NH) values corresponds to the sky--plane projection
  of a spherical exponential distribution of atomic volume density,
  with the indicated central log(\NH) and scale height in arcsec. The
  asymmetric profiles of CHVC100$-$49$-$383 and CHVC148$-$32$-$144 show
  large departures from the overlaid curves. In the panel for
  CHVC158$-$39$-$285, the accumulated fractional flux (assuming
  circular source symmetry) is indicated by the increasing curve which
  is labeled to the right. }
\label{fig:nh1}
\end{figure*}

\begin{figure*}
\resizebox{14cm}{!}{\includegraphics{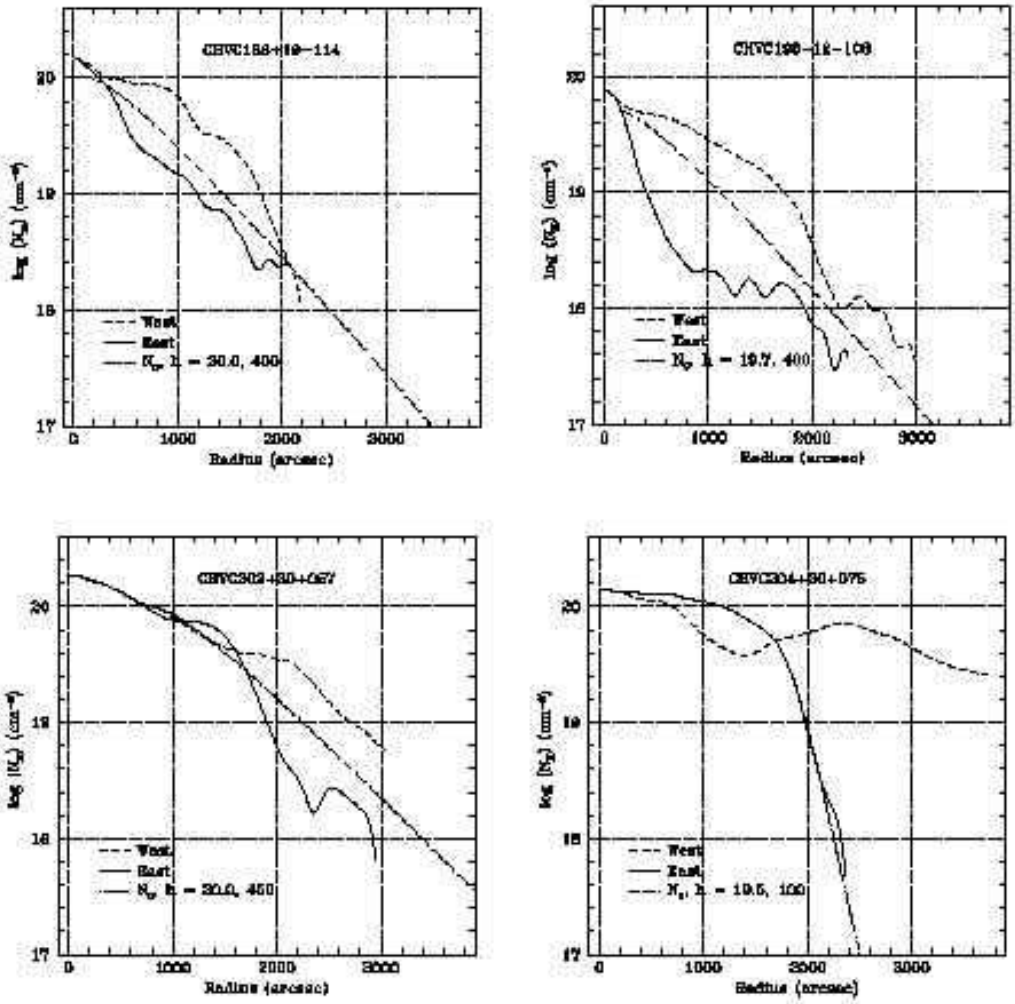}}
\caption{
  Column density profiles of the indicated CHVCs as in
  Fig.~\ref{fig:nh1}. } 
\label{fig:nh2}
\end{figure*}

The plots in Figs.~\ref{fig:nh1} and \ref{fig:nh2} show the variation
of \NH~with radius measured from the position of peak column density in
the single deeper--integration cross--cut made across each of the CHVC
targets.  The {\sc rms} sensitivities (measured over 10 \kms) of these
slices are indicated in column 4 of Table \ref{tab:sample}; the total
velocity extents over which the column depths were determined are
indicated in column 5 of the table.  The plots of \NH~against radius
were truncated at a level of 2 times the {\sc rms} value.  The Eastern
and Western halves of the CHVCs are plotted separately in the panels of
Fig.~\ref{fig:nh1} and \ref{fig:nh2}, in order to allow assessment of
the reflection--symmetry seen in the two halfs of the cross--cut. Since
the choice of origin is arbitrary, it is the profile slope at a given
column density which should be compared rather than simply the degree
of overlap with this particular choice of origin. The dotted curve
overlaid on each of the panels is described below.

\begin{figure*}
\resizebox{16cm}{!}{\includegraphics{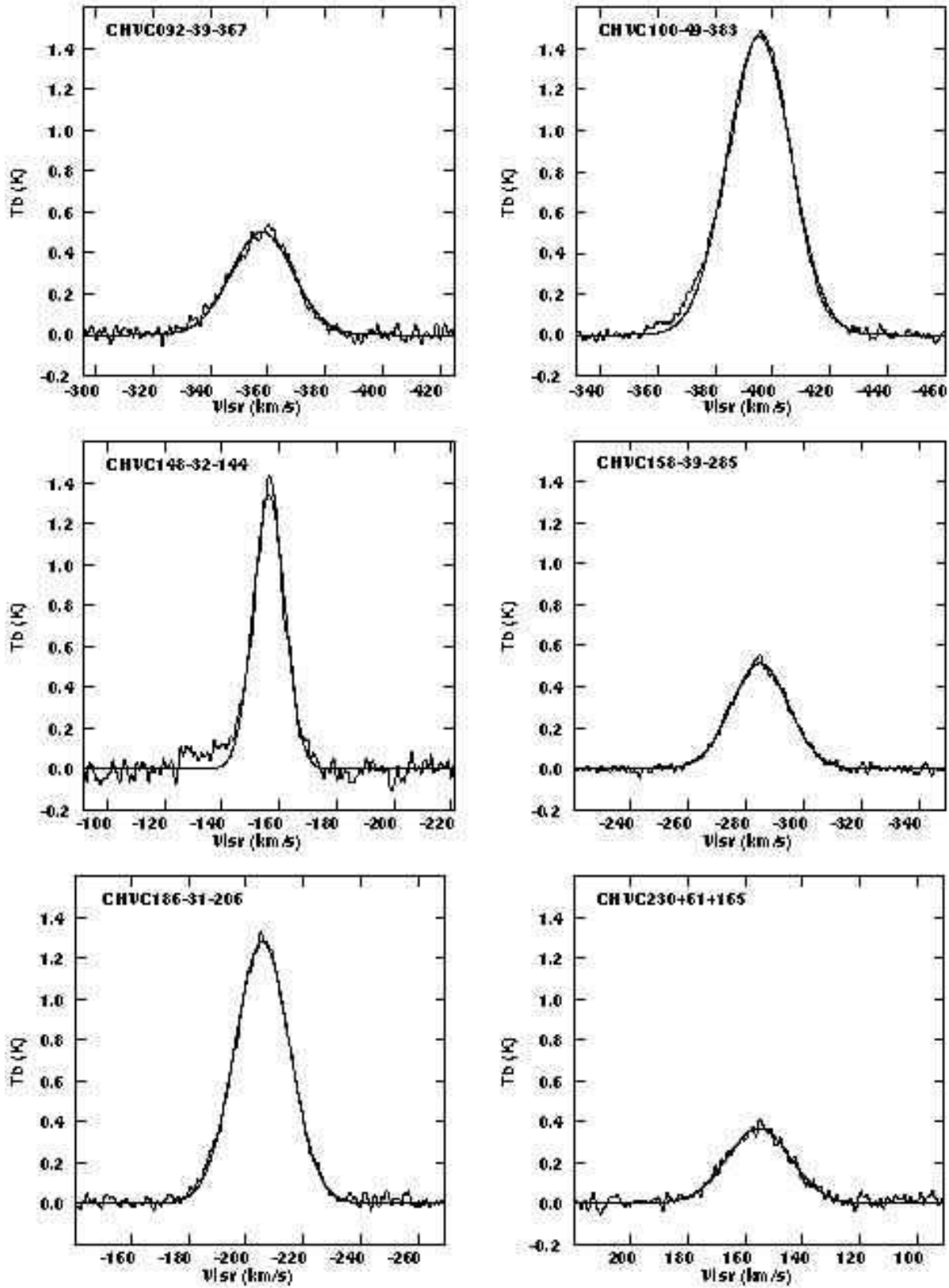}}
\caption{
  Spectra of the emission peaks in the indicated CHVCs.  Each
  spectrum samples the emission peak in the lower--left panel of
  Figs.~\ref{fig:h092}--\ref{fig:h230}. Overlaid on each spectrum is
  the fit of a single Gaussian: the \vlsr, $T_{\rm max}$, and FWHM
  corresponding to this single Gaussian are tabulated in Table
  \ref{tab:results}. }
\label{fig:spec1}
\end{figure*}

\begin{figure*}
\resizebox{16cm}{!}{\includegraphics{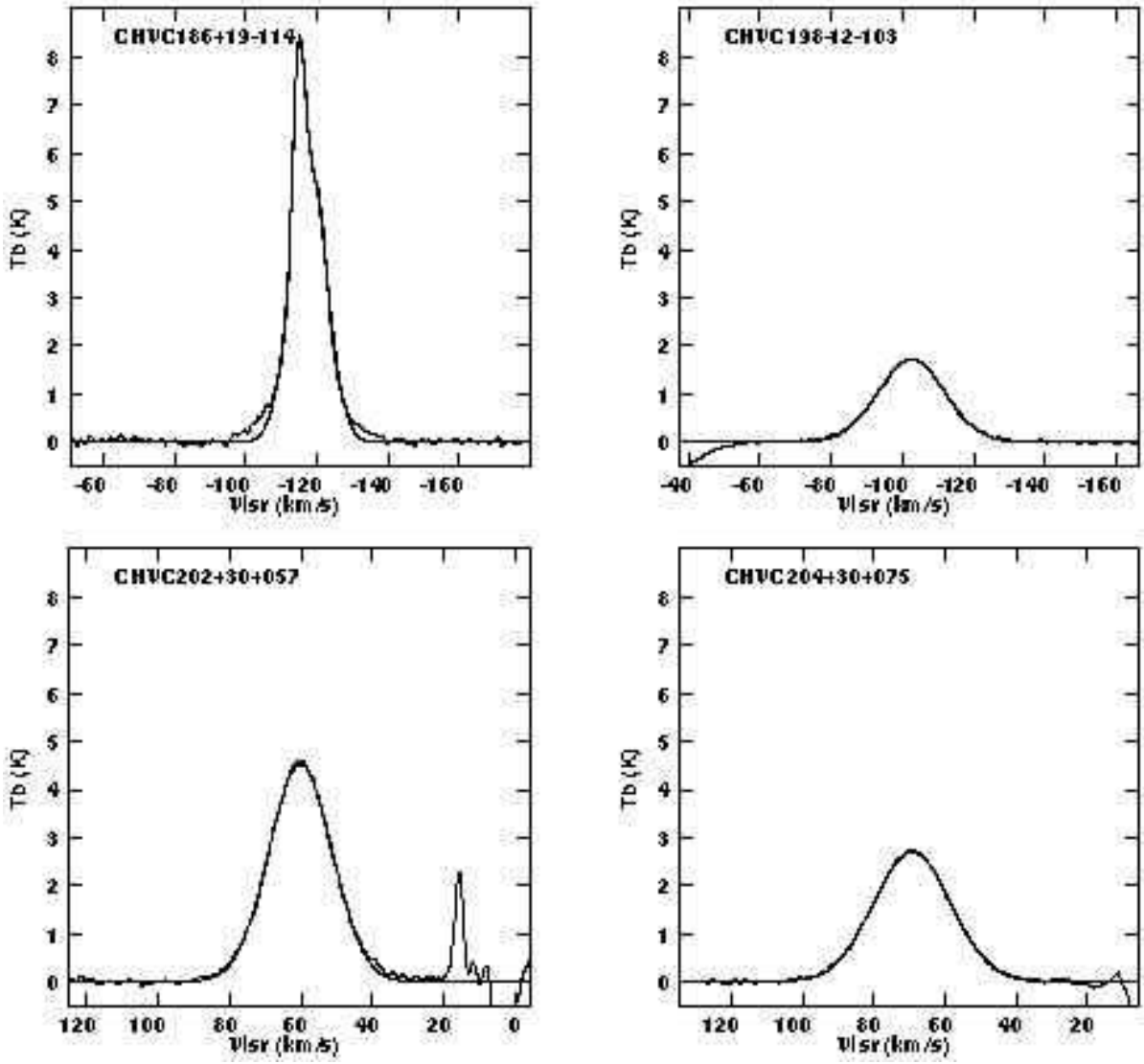}}
\caption{
  Spectra of the emission peaks in the indicated CHVCs, as in
  Fig.~\ref{fig:spec1}. A single Gaussian resulted in an adequate fit
  for three of the objects, but not for CHVC\,186+19$-$114, where
  the shape of the spectral cross--cut requires (at least) two
  components.  The corresponding values of \vlsr, $T_{\rm max}$, and
  FWHM are tabulated in Table \ref{tab:results}.  }
\label{fig:spec2}
\end{figure*}

The plots in Figs.~\ref{fig:spec1} and \ref{fig:spec2} show the $T_{\rm
B}$,\,\vlsr~spectrum measured in the direction of the column density
peak identified in the long--integration driftscans shown in the panels
on the lower left of Figs.~\ref{fig:h092}--\ref{fig:h230}.  Overlaid on
each spectrum (at 1~\kms~velocity resolution) is the profile
representing a Gaussian fit. A single Gaussian was fit to all of the
profiles expect for the one observed for CHVC\,186+19$-$114, for which
two Gaussians were judged inevitable.  Note that the information for
each CHVC represented in this figure refers only to a single
line--of--sight toward one peak, namely the peak of the emission
centroid chosen for the cross--cut.

\setlength{\tabcolsep}{4pt}
\begin{flushleft} 
\begin{table*} 
\caption[]{CHVC properties. Column 1 gives the object
designation. Columns 2, 3, and 4 give the $v_{\rm LSR}$, the peak
temperature, and the velocity FWHM, respectively, determined from the
Gaussian fits (plotted in Figs.~\ref{fig:spec1} and \ref{fig:spec2}) to
the emission peaks of the deep constant--declination driftscans. Column
5 gives the logarithm of the \hi column density in the direction of the
emission peak, integrated over the velocity ranges limited by the
vertical lines in the panels on the lower left of
Figs.~\ref{fig:h092}--\ref{fig:h230}.  Columns 6 and 7 give the total
WNM atomic column density and exponential scale--length, respectively,
derived as discussed in the text under the assumption of approximate
spherical symmetry.  Column 8 gives the distance calulated from
eqn.\ref{eqn:Dis} and a nominal CNM/WNM transition pressure. Column 9
gives the distance calulated assuming a mean outer disk scale-length,
$h_B$=10.6 kpc, as found for nearby low mass dwarf galaxies. }
\label{tab:results}
\begin{tabular}{crccccccc} 
\noalign{\smallskip} \hline 
Name & $v_{\rm LSR}~~ $ & $T_{\rm max}$& {\small FWHM\,} & log(\NH) 
& log(\NH(0)) & $h_B$ & Dist.\,(eqn.\,\ref{eqn:Dis}) &
Dist.\,($h_B$=10.6 kpc) \\ 
{\small CHVC}\,$lll \pm bb \pm vvv$ &(\kms) & (K) & (\kms) &
Gauss--fit  & halo--fit & (arcsec)  & (kpc) & (kpc) \\
\noalign{\smallskip} \hline 
CHVC\,092$-$39$-$367 & $-$358~~ & 0.5  & 25 & 19.4 & 19.4 & 300 &
280 & 730 \\ 
CHVC\,100$-$49$-$383 & $-$395~~ & 1.5  & 26 & 19.9 & 19.4 & 350 & 
--- & --- \\
CHVC\,148$-$32$-$144 & $-$156~~ & 1.3  & 13 & 19.6 & 19.3 & 250 & 
--- & --- \\ 
CHVC\,158$-$39$-$285 & $-$285~~ & 0.5  & 23 & 19.4 & 19.4 & 550 &
150 & 400 \\ 
CHVC\,186$-$31$-$206 & $-$206~~ & 1.3  & 22 & 19.7 & 19.6 & 500 &
270 & 440 \\ 
CHVC\,186+19$-$114   & $-$118~~ & 5.9  & 12 & 20.1 & 20.0 & 400 &
840 & 550 \\
                     & $-$115~~ & 3.4  & 3.5& 19.4 &      &     &
    &     \\
CHVC\,198$-$12$-$103 & $-$102~~ & 1.7  & 22 & 19.9 & 19.7 & 400 &
420  & 550 \\
CHVC\,202+30+057     & +60~~    & 4.5  & 21 & 20.3 & 20.0 & 450 &
740  & 490 \\
CHVC\,204+30+075     & +69~~    & 2.7  & 26 & 20.1 & 19.5 & 100 &
---  & ---  \\ 
CHVC\,230+61+165     & +155~~   & 0.4  & 26 & 19.3 & 19.2 & 350 &
150 & 320 \\
\noalign{\smallskip} \hline 
\end{tabular} 
\end{table*} 
\end{flushleft} 
 
\section{Results for the selected CHVCs}
\label{sec:results}

We briefly summarize the structural and kinematic data obtained for
each of the ten observed fields.  We first comment on the appearance of
each object in the shallow Nyquist--sampled $1\deg \times 1\deg$ grids
and in the deeper integrations along the two--degree cross--cut, as
shown in the respective panels of Figs.~\ref{fig:h092}--\ref{fig:h230},
and then comment on the spatial and kinematic properties of one of the
principal cores of each CHVC, as shown in the respective panels
of Figs. \ref{fig:nh1} and \ref{fig:nh2} and Figs.
\ref{fig:spec1} and \ref{fig:spec2}.  Some of the results are summarized in
Table~\ref{tab:results}. 
 
\subsection{CHVC\,092$-$39$-$367}

CHVC\,092$-$39$-$367 is one of the weaker objects in the BB99
catalog, but one at a relatively extreme negative radial velocity.
With the $3\farcm5$ angular resolution of the Arecibo telescope, the
feature shows a complex core/halo morphology.  The two--dimensional
shallow--mapping image, shown in the upper left of Fig. \ref{fig:h092},
reveals an ensemble of some half dozen separate cores, largely embedded
in extended emission.  The velocity centroids of the individual cores
range between \vlsr\,$=-350$ and $-370$ \kms. The $1\deg \times 1\deg$
image in the upper--left panel shows no well--defined structural axis,
and the intensity--weighted velocity field of this two--dimensional
image, shown in the panel on the upper--right of Fig.~\ref{fig:h092},
also shows no well--defined kinematic pattern.

The longer--integration driftscan cross--cut spanned two degrees of
right ascension at $\delta = 17\deg24\arcmin$.  This declination was
chosen because it crosses the core with the highest column density; but
it will be clear that a cross--cut at a different constant declination,
i.e.  crossing a different region of the CHVC, might reveal different
details.  The cross--cut, displayed as an $\alpha$,\,\vlsr~map in the
lower left of Fig.~\ref{fig:h092}, shows two of the brighter cores
evident in the moment--map image, as well as one of the fainter cores
evident in the two--dimensional image, but also an additional minor
core feature which is out of the field of view of the $1\deg \times
1\deg$ image.  With the higher sensitivity of the longer integration,
all four of these cores are revealed to be embedded in a common
envelope.

The three panels grouped on the lower right of Fig.~\ref{fig:h092} show
the \vlsr, the velocity FWHM, and the logarithm of \NH, as measured
along the constant--declination cross--cut. The data show an overall
kinematic gradient spanning about 20 \kms over the 1.5 degree extent of
the entire source. Superposed on this global gradient are the much more
abrupt gradients associated with the individual compact cores. These
abrupt gradients reach magnitudes of 10--15 \kms on scales of only
5--10~arcmin.

The FWHM velocity width varies between about 25 \kms, in unconfused
regions, and 40 \kms, in those regions where different components
overlap along the line-of-sight.  (We comment below on the
interpretation of these FWHM values, which, of course, represent upper
limits to the {\it in situ} kinematics.)  The column depths of the two
principal cores are about $10^{19.3}$ cm$^{-2}$; the diffuse envelope
in which the cores are embedded has been traced to \NH~levels of about
$10^{18.0}$ cm$^{-2}$.

The panel on the upper left in Fig.~\ref{fig:nh1} shows the variation
of log(\NH) with angular distance from the peak column density,
separately for the Western and Eastern halves of the object.  The
column depths can be traced on both halves to levels somewhat less than
$10^{18}$ cm$^{-2}$, at an angular distance of some $45\arcmin$ from
the emission centroid of the constant--$\delta$ slice.  The panel on
the upper left in Fig.\ref{fig:spec1} represents a $T_{\rm
B}$,\,\vlsr~spectrum through the emission centroid.  Overlaid on the spectrum
is a Gaussian distribution corresponding to the \vlsr, $T_{\rm
  max}$, and FWHM parameters listed in Table~\ref{tab:results}.

\subsection{CHVC\,100$-$49$-$383}

C{\sc HVC}\,100$-$49$-$383, like the previous object also at an
exceptionally extreme velocity, appears in the BB99 catalog as a
particularly simple object.  The confirming data sought by BB99 had
involved additional data from the Dwingeloo 25--m telescope,
Nyquist--sampled on a $15\arcmin$ grid. The Arecibo $1\deg \times 1\deg$
mapping image shown in the upper left of Fig.~\ref{fig:h100} shows
considerably more detail at this higher resolution and greater sensitivity:
the brightest core is offset to the North-East of the source centroid,
while secondary cores are distributed over a wider region and are all
embedded in a common envelope.

The deeper cross--cut shown in the lower left of Fig.~\ref{fig:h100}
scans the $2\deg$ strip along the declination, $11\deg19\arcmin$, where
the brightest core is most intense. There is only a modest velocity
gradient along this cut, amounting to less than 10 \kms.  However,
within the bright core there are rapid velocity reversals on scales of
5--10 arcmin.  The velocity FWHM varies between about 25 and 35 \kms.
The core peaks at a column depth of $10^{19.9}$ cm$^{-2}$.  The
envelope could be followed on the lower--$\alpha$ side to \NH~$\sim
10^{18.3}$ cm$^{-2}$. 

A noteworthy aspect of the morphology of this CHVC is that the
off-center location of the brightest core component gives it the
appearance of being more sharply bounded on one side than the
other. The panel on the upper right of Fig.~\ref{fig:nh1} shows the
column density profiles to the East and West of the position of the
bright core. While the Eastern profile shows a rapid decrease in \NH\ to
values below $10^{18}$ cm$^{-2}$, corresponding to the actual edge of
the source, we detect emission in excess of $10^{18}$ cm$^{-2}$ out to
the limit of our coverage in the West. Only by extending the coverage
significantly further to the West would it become clear whether the WNM
halo in this source is itself symmetric or not. Similar off-center
locations of bright cores are seen in several of the other
CHVCs described here, and are discussed further below. The spectrum
corresponding to the peak of the deep driftscan and plotted on the
upper right of Fig.~\ref{fig:spec1} indicates that the CNM core shows 
substantial kinematic symmetry.

\subsection{CHVC\,148$-$32$-$144}

C{\sc HVC}\,148$-$32$-$144 appears in the BB99 catalog as a simple, but
somewhat elongated object, at a modest deviation velocity.  The Arecibo
$1\deg \times 1\deg$ shallow image shown in the upper left of
Fig.~\ref{fig:h148} does not fully encompass the CHVC; two prominent
cores are evident in the region mapped, each having --- as shown in the
intensity--weighted velocity field image of the upper right panel ---
its own characteristic velocity.  The cores are enclosed in a common
envelope.  The constant declination for the deeper $2\deg$ cross--cut
was chosen at $\delta = 26\deg22\arcmin$, near one of the peaks in the
shallow image.

The deep constant--declination driftscan through the CHVC is shown in
the lower left of Fig.~\ref{fig:h148}.  In this direction, two
intensity peaks are seen, the principal one centered near $\alpha =
2^{\rm h}26^{\rm m}40^{\rm s}$ and \vlsr\,$=-155$ \kms, and a secondary
one near $\alpha = 2^{\rm h}23^{\rm m}30^{\rm s}$ and \vlsr\,$=-140$
\kms.  It is plausible that the rather large velocity FWHM of 35
\kms~tabulated for this object by BB99 on the basis of
mapping with the Dwingeloo 25--m telescope refers to the accumulated
kinematics contributed by several cores, each at a somewhat different
centroid velocity.  Thus the principal core measured in the Arecibo
data has a FWHM of 17 \kms, whereas the secondary core has a FWHM of
about 25 \kms.

The cross--cut slice through CHVC\,148$-$32$-$144 shows that this
object also is characterised by a spatial offset of the brighter core
from the centroid of the underlying halo. Only on the Eastern side of
the source does the coverage extend far enough to adequately sample the
edge where it can be followed to column depths as low as some
$10^{17.7}$ cm$^{-2}$, at $\sim6\arcmin$ from the bright core.  The
middle panel on the left of Fig.~\ref{fig:spec1} shows a
single--Gaussian fit to the spectral cut through the \sc CHVC\rm.  The
residuals from the single--Gaussian fit are substantially greater than
the noise level, and are systematic in nature: evidently this CHVC has
spatial structure which is essentially unresolved in angular extent at
the limit of the Arecibo data.

\subsection{CHVC\,158$-$39$-$285}

CHVC\,158$-$39$-$285, at a substantial negative velocity, is one of the
faintest in the catalog of BB99.  The $1\deg \times 1\deg$
two--dimensional Arecibo image shown in the upper left of
Fig.~\ref{fig:h158} reveals a core of higher column density \hi emission,
surrounded by additional sub-structure.  The velocity
field shown in the upper right of Fig.~\ref{fig:h158} shows a
systematic kinematic gradient oriented East-West along the long axis of
the brightest portion of the object.

The more sensitive observations constituting the $2\deg$
$\alpha$,\,\vlsr~cross--cut through the core were made at $\delta =
16\deg17\arcmin30\arcsec$.  The \NH~value reaches some $10^{19.4}$
cm$^{-2}$ at the peak of the core.  \NH~in the diffuse envelope can be
traced to about $10^{17.6}$ cm$^{-2}$ on the Western side, where the
centroid velocity is some 20 \kms~different from the velocity
characteristic of the peak of the core emission. The spatial coverage
to the East is not sufficient to reach the edge of the envelope.

The position,\,velocity cross--cut reveals a clear kinematic gradient
within the high column density core of this source, spanning a velocity
difference of 30 \kms, and possibly suggesting rotation in a flattened
disk.  Because of its interesting kinematic structure and limited
angular extent, this core is a good candidate for synthesis mapping.

The panel on the middle right of Fig.~\ref{fig:nh1} shows the \NH\
profiles to the East and West of the position of peak column
density. The two edge profiles are remarkably symmetric over the
measured range of column density.  The panel on the middle right of
Fig.~\ref{fig:spec1} shows that a single Gaussian accounts for much of
the emission.

\subsection{CHVC\,186$-$31$-$206}

CHVC\,186$-$31$-$206, not previously identified as a high--velocity
cloud, shows an elongated structure in the two--dimensional shallow
image on the upper left of Fig.~\ref{fig:a186}, with one principal core
and at least two secondary ones, enclosed in a common envelope.  The
velocity field shown on the upper right of the figure shows that the
different core substructures of this CHVC occur at different
characteristic velocities, spanning some 12 \kms. The velocity gradient
oriented along the elongated axes of the feature might suggest 
rotation.

The declination for the longer--integration cross--cut was chosen to
coincide with the peak of the CHVC flux, at
$6\deg36\arcmin19\arcsec$.  The emission along this slice peaks at a
velocity of $-206$ \kms.  The envelope can be followed to \NH~levels of
somewhat less than $10^{18}$ cm$^{-2}$.

The panel shown on the lower left of Fig.~\ref{fig:nh1} indicates that
while the Eastern edge of this CHVC has been delineated to below
\NH=$10^{18}$ cm$^{-2}$, the coverage has not been sufficient to fully
characterize the Western edge.  The fit by a single Gaussian shown in
the corresponding panel of Fig.~\ref{fig:spec1} shows residuals of
greater amplitude than expected from the noise figure of the data: this
object might have structure which remains unresolved in
angle by the Arecibo beam.  (The results obtained for the few CHVCs
which have been observed both at Arecibo and with the WSRT show that it
is emminently plausible to expect such unresolved structure at the
Arecibo resolution.)

\subsection{CHVC\,186+19$-$114}

CHVC\,186+19$-$114 is one of the brighter
($T_{\rm max} = 1.03$\,K in the LDS) objects in the CHVC tabulation of
BB99.  The shallow Arecibo image shows an elongated structure,
extending beyond the limits of the $1\deg \times 1\deg$ map shown on
the upper left of Fig.~\ref{fig:h186}.  The panel on the upper right of this
figure shows that the centroid velocity on the lower--$\delta$ side is
some 10 \kms~less extreme than on the higher--$\delta$ side.

The declination for the $2\deg$ longer--integration cross--cut was
chosen to coincide with the peak column density, at $31\deg52\arcmin$.
The stucture revealed by the observations along this slice is that of
an elongated core.  But the emission sampled in the constant--$\delta$
cut is probably not contributed by a single, simple core. The panel in
the upper left of Fig.~\ref{fig:nh2} shows the two opposing \NH\
profiles which can be traced only down to about $10^{18}$ cm$^{-2}$.
The corresponding panel in Fig.~\ref{fig:spec2} shows that the
kinematics measured through the core can not be fit by a single
Gaussian, but requires (at least) two features to account for the shape
of the spectrum.  The emission on the wings of the spectrum remaining
unaccounted for by a two--component fit suggests that the core/halo
morphology which seems characteristic of CHVCs imaged at high
resolution by the WSRT pertains here too, with the Arecibo resolution
only marginally sufficient to resolve the core/halo structure.

\subsection{CHVC\,198$-$12$-$103}

CHVC\,198$-$12$-$103 appears in the BB99
catalog as a moderately bright ($T_{\rm max}= 0.48$\,K), broad feature
with a rather simple form.  This impression remains under more detailed
scrutiny with the Arecibo telescope.  The $1\deg \times 1\deg$ imaging
data shown on the upper left of Fig.~\ref{fig:h198} reveals a
highly asymmetric core component with some internal sub-structure. The
core is characterized by a much sharper gradient to the East than to
the West. The intensity--weighted velocity centroids of the two
principal emission peaks, shown in the panel on the upper right of the
figure, differ by only a few \kms.

The longer--integration $2\deg$ cross--cut, shown in the lower left
panel of Fig.~\ref{fig:h198}, sliced the object at declination
$7\deg54\arcmin$, approximately through the central location of the
feature.  The $\alpha$,\,\vlsr~slice shows both the strong asymmetry of
the high \NH\ core as well as the much more symmetric halo in which the
core is embedded. The object as a whole displays a kinematic gradient,
spanning \vlsr~from $-102$ \kms~to $-110$ \kms.

The lopsided nature of the high \NH\ core in this CHVC is evident in
the upper right panel of Fig.~\ref{fig:nh2}, showing the edge profiles.
The Eastern edge of the core is poorly resolved at the $3\farcm5$
resolution of the Arecibo data.  However, it is remarkable that the low
\NH\ envelope of this core (below about $10^{18.2}$ cm$^{-2}$) shows a
high degree of symmetry.  The kinematic structure, shown in the upper
right of Fig.~\ref{fig:spec2}, is well fit by a single Gaussian, with
FWHM of 22 \kms.  The object is clearly defined against its spatial and
kinematic background, and so could be measured to column depths as low
as $10^{17.5}$ cm$^{-2}$.

\subsection{CHVC\,202+30+057}

C{\sc HVC}\,202+30+057 is one of the first objects encountered at
positive {\sc lsr} velocities, as longitudes increase beyond that of
the Galactic anticenter.  It is found in the general
$(l,\,b,\,$\vlsr) region populated by the anomalous--velocity features
studied first by Wannier \& Wrixon (\cite{wawx72}) and Wannier et al.
(\cite{wann72}).  The two--dimensional shallow Arecibo image in the
upper left of Fig.~\ref{fig:h202} shows a rather simple elongated
structure, blending with \hi emission from the conventional Milky Way
disk at the higher declinations.

The declination chosen for the deep $2\deg$ cross--cut,
$21\deg46\arcmin$, slices the peak \NH\  concentration.  The
$\alpha$,\,\vlsr~slice shows that the principal core can be adequately
separated from Milky Way contamination.  The feature is quite broad,
with a velocity FWHM of about 25 \kms, and little gradient over the
region contributing most of the emission.

The \NH\ profile for this object, plotted on the lower left of
Fig.~\ref{fig:nh2}, reaches column densities below $10^{18}$ cm$^{-2}$
toward the East, but the spatial coverage to the West is insufficient
to delineate the true source extent. The kinematic behavior shown in
Fig.~\ref{fig:spec2} is rather straightforward.  Although the
single--Gaussian fit leaves a substantial residual on the low--velocity
wing, it is not clear if this residual is due to contamination by gas
in the disk of the Milky Way rather than to a broad halo of diffuse gas
in the CHVC.

\subsection{CHVC\,204+30+075}

CHVC\,204+30+075 was cataloged by BB99 and subsequently was one of the
six CHVCs imaged with the Westerbork Synthesis Radio Telescope by BB00.
In the data from the Dwingeloo 25--m telescope the object is quite
intense, with a peak temperature of 1.19 K, and the largest LDS flux,
305 Jy\,\kms, of any of the cataloged CHVCs.  The Arecibo
two--dimensional mapping is shown in the upper left of
Fig.~\ref{fig:h204}: for this target, the Nyquist--sampled grid was
extended in size to $1\fdg5 \times 1\deg$ in order to accommodate
essentially the entire CHVC within the boundaries of the accessible
emission.  (Note that the two contours near ($08^{\rm h}26^{\rm m}$,
$20\deg10\arcmin$) are a local minimum in \NH.)  The Arecibo image
shows two principal cores, in a common envelope. The distribution of
the intensity--weighted velocity centroids shown in the panel on the
upper right of Fig.~\ref{fig:h204} shows substantial structure, with
the Northern core showing a particularly pronounced kinematic gradient.

The constant--declination deep--integration slice sampled a strip along
$\delta = 20\deg1\arcmin30\arcsec$.  The \hi data on the strip allow
tracking of the Eastern edge of the object down to about
\NH~=~$10^{18}$ cm$^{-2}$, but the spatial coverage was not sufficient
to delineate the Western edge beyond \NH~=~$10^{19.4}$ cm$^{-2}$.

The spatial and kinematic cuts through the centroid of this CHVC are
shown in the panels on the lower right of Figs.~\ref{fig:nh2} and
\ref{fig:spec2}, respectively.  The Eastern \NH\ profile is only
marginally resolved with the $3\farcm5$ Arecibo beam.  A single
Gaussian of FWHM 26 \kms~accounts for the spectral cross--cut through
the centroid.

\subsection{CHVC\,230+61+165}

CHVC\,230+61+165 was cataloged by BB99 and was also
one of the six CHVCs imaged with the WSRT by BB00.  The primitive map
of this object shown by BB99 was based on LDS data sampled on a
$0\fdg5$ lattice; on such a coarse grid the core was essentially
unresolved, but the map did show a hint of some surrounding emission.
The shallow Nyquist--sampled Arecibo image shown in the upper left of
Fig.~\ref{fig:h230} reveals several resolved cores, clearly embedded in
a common envelope.  The panel on the upper right shows that the several
cores are each characterized by a somewhat different velocity centroid.

The longer--integration $\alpha$,\,\vlsr~slice shown at the lower left
of Fig.~\ref{fig:h230} was made along declination $15\deg28\arcmin$, a
direction crossing the CHVC near the centroid of the emission
from the object as a whole, but passing through one of the minor cores.
The position,\,velocity map shows a knot of emission, centered near 156
\kms, with little variation in either centroid velocity or in velocity
FWHM.  

The panels on the lower right of Figs.~\ref{fig:nh1} and
\ref{fig:spec1} show the spatial and kinematic cross sections,
respectively, through the centroid of the emission sampled in the
longer--integration slice.  Although the halo may be detected below
\NH~=~$10^{18}$ cm$^{-2}$ further to the East than the West this is
very near the noise floor for this field.  The single--Gaussian fit,
although roughly adequate in shape, leaves residuals above the noise
level which are suggestive of unresolved detailed structure.

\section{Insights based on comparison of the Arecibo and 
WSRT results for two CHVCs}
\label{sec:comp}

It is instructive to compare the properties of CHVCs revealed by the
Arecibo filled--aperture telescope with those revealed by the
Westerbork synthesis instrument.  We anticipate in this
section the general conclusion that neither a large filled--aperture
antenna such as the Arecibo one nor a synthesis instrument such as the
WSRT will, alone, suffice to reveal the details of the core/halo
morphology which pertains to the CHVCs. The picture which has emerged from
the total of eight CHVCs imaged with the WSRT by Wakker \& Schwarz
(\cite{wakk91b}) and by BB00 is that of compact cores with angular sizes
typically of a few arcmin.  The \hi linewidths of the cores are rather
narrow (usually less than 10
\kms~FWHM) and they often display significant velocity gradients along
the long dimension of an elliptical extent.  At resolutions coarser
than the angular size of the cores, such cores will, of course, remain
unresolved.  

CHVC\,204+30+075 and CHVC\,230+61+165 have now been observed at both the
WSRT and at Arecibo. The CNM cores of these objects were imaged by BB00
using the WSRT at 28 arcsec angular resolution; the 3.5--arcmin
resolution of the Arecibo data is insufficient to resolve the cold knots.
On the other hand, the interferometer does not detect the diffuse halos
which so clearly envelop the objects studied in the Arecibo data.
More important is the complete change of character of the
\hi line profiles when full sensitivity to the diffuse halos is
present, as in the Arecibo total--power data presented here. As seen in
the figures, as well as in Table~\ref{tab:results}, most
lines--of--sight are dominated by the emission from the diffuse halos,
leading to much broader total \hi linewidths of about 25~\kms~FWHM. The
velocity gradients of individual cores are diluted by this background
contribution to such an extent that they can often not even be
discerned.

\begin{figure*}[h]
\resizebox{18cm}{!}{\includegraphics{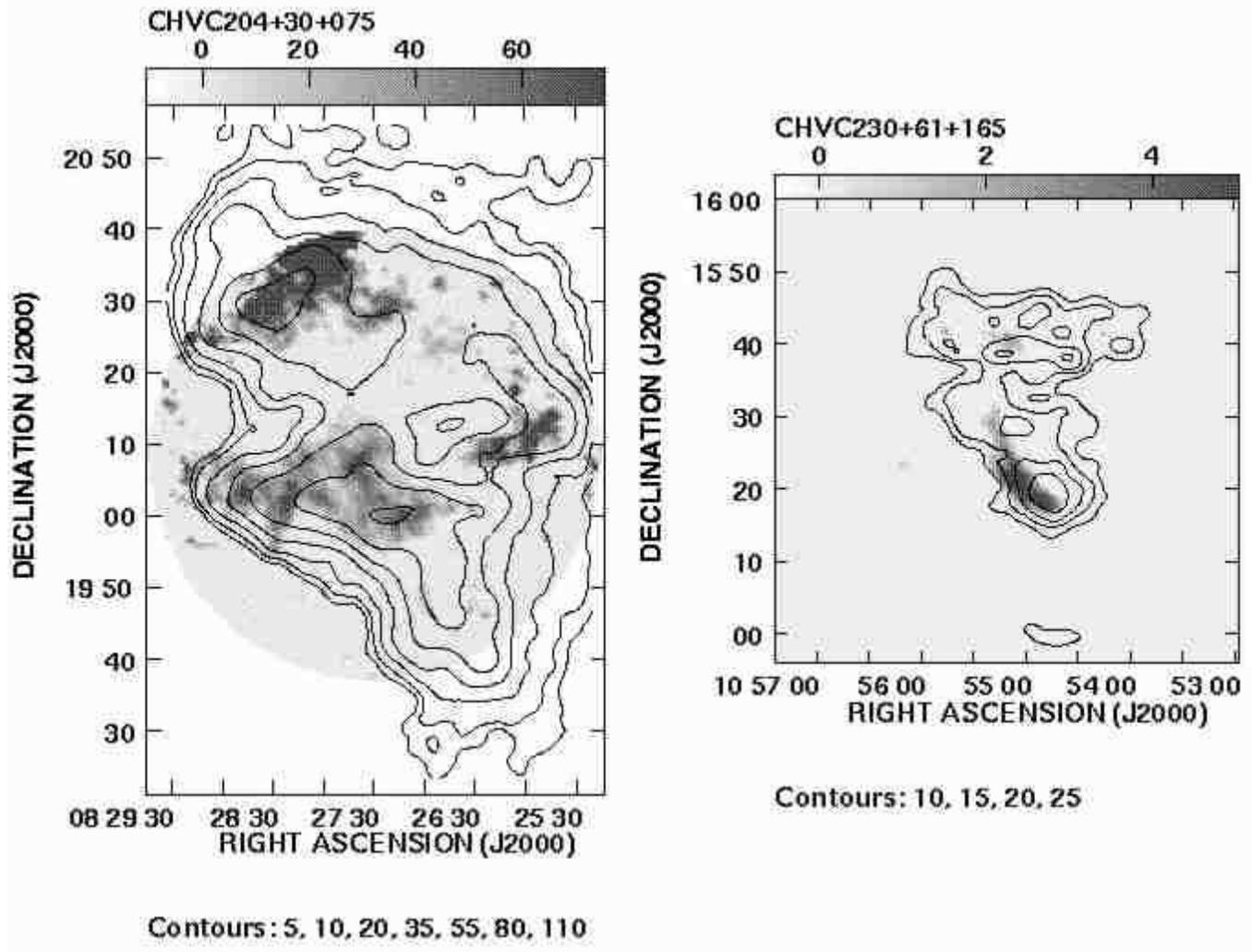}}
\caption{
  Overlay of WSRT and Arecibo \NH~data for CHVC\,204+30+075 and
  230+61+165. The WSRT detected \hi column density at 1~arcmin (for
  CHVC\,204+30+075) and 2~arcmin (for CHVC\,230+61+165)
  resolution is indicated by the grey--scale background. The Arecibo
\NH~contours from Figs.~\ref{fig:h204} and \ref{fig:h230} are overlaid.  } 
\label{fig:h204+230c}
\end{figure*}

The left--hand panel of Fig.~\ref{fig:h204+230c} shows the \hi~column
density distribution in CHVC\,204+30+075 derived from the WSRT data at
an angular resolution of 1 arcmin (from BB00) overlaid on the Arecibo
data.  Effectively none of the diffuse emission is detected in the WSRT
data.  Consequently, the velocity field and spectra shown in Fig.~8 of
BB00 are dramatically different from those shown in
Fig.~\ref{fig:h204}. To affirm that the WSRT \NH~results are compatible
with those from Arecibo requires the realization that the WSRT has not
responded to the diffuse halo prominently seen enveloping the cores in
the Arecibo data, whereas the Arecibo angular resolution has not been
sufficient to reveal the core details. The comparison of the column
density distribution of \hi in CHVC\,230+61+165 made with the WSRT and
Arecibo telescopes, shown in the right--hand panel of
Fig.~\ref{fig:h204+230c}, leads to similar conclusions. Only the
compact CNM cores, with their narrow emission lines, are detected in
the synthesis data; the total power data, on the other hand, are
completely dominated by the diffuse, broader--linewidth, WNM halos.

This same effect has substantial implications for the interpretation of
\hi observations of some nearby spiral galaxies. Dickey et
al. (\cite{dick90}) and Rownd et al. (\cite{rown94}) have measured the
radial distribution of \hi linewidths in the galaxies NGC~1068 and
NGC~5474 from VLA imaging data, and quote velocity dispersions as low
as 6~\kms~(or 14~\kms~FWHM) in the outer disks of these systems (at \NH
$\sim10^{20}$ cm$^{-2}$). However, the VLA observations only detected
about half of the total \hi flux in these galaxies. It is clear that
the missing flux is in a smoothly distributed component --- hence its
non--detection in the synthesis data --- and quite likely that it
represents a WNM component with the broader intrinsic linewidth (of
about 25~\kms~FWHM) we have measured here. This conjecture is supported
by the Arecibo observations of M33 of Corbelli et al. (\cite{corb89}),
who consistently find \hi linewidths of about 25~\kms~FWHM (whenever
the signal--to--noise ratio was adequate) at the largest detected radii
of that galaxy.  The generality of this conclusion depends, of course,
on the match of interferometer sensitivity to diffuse structures in the
specific object under study. From our resolved detection of the warm
halos in the Arecibo data we can make concrete estimates of the
required brightness sensitivity. Peak brightnesses of the WNM halos
seen in Figs.~\ref{fig:h092}--\ref{fig:h230} are between 20 and 50 mK
over about 20~\kms.  Direct WNM detection therefore requires a
brightness temperature {\sc RMS}, $d$T$_{B}~<$~10~mK over 20~\kms for
emission that fills the beam. For comparison, a 12 hour integration with
the VLA D configuration provides a theoretical {\sc RMS} of 60~mK over
20~\kms\ in a naturally weighted image (Perley \cite{perl00}) for which
the beamsize is about 65~arcsec {\sc FWHM}. If a physical resolution of
3~kpc were required, this beamsize would suffice out to a distance of
10~Mpc. However, the 60~mK {\sc RMS} of a 12 hour observation would
only ever allow detection of the CNM peaks and the very brightest
portions of an underlying WNM.

The WSRT data have indicated that the CHVC cores do not have an
intrinsically Gaussian spatial or kinematic form, either when viewed
individually or as an ensemble of several cores within one CHVC halo.
The accuracy of a single--Gaussian fit to the Arecibo cross--cuts is
not at odds with this conclusion, if one recognizes that it is
predominantly the flux from the halo which is being fit.  If the
cross--cut emission were being contributed by a collection of
narrow--linewidth CNM cores, kinematically spanning some 20 \kms, then
the cross--cut spectra would be characterized by the same FWHM as has
been observed, but the wings of the spectrum would be steep; in fact,
the spectral wings are consistent with WNM linewidths.

Furthermore, the conclusion which we have been able to draw from the
Arecibo observations reported here, namely that the column depths in
the outer regions of the CHVCs fall off as an exponential with radius,
implies that for sensitivity--limited data, both the measured sizes and
fluxes will depend on the sensitivity and resolution employed.  (We
note in this regard that it is reasonable to expect that observations,
made with currently available instrumental parameters, of any analogous
objects which might be located beyond the Local Group would be severely
sensitivity limited.)  To illustrate this point with a concrete
example, we have plotted the cumulated fractional flux as function of
radius in Fig.~\ref{fig:nh1} for CHVC\,158$-$39$-$285, assuming that the
measured radial profile for this object has azimuthal symmetry.
Resolved observations of this object (with an angular resolution at
least as good as 500~arcsec) with a limiting column density sensitivity
of $\sim10^{19}$cm$^{-2}$ would measure a source radius of about 1000
arcsec and only detect some 50\% of the total flux density. Only with
a resolved column density sensitivity of $\sim10^{18}$cm$^{-2}$ or
better would more than 90\% of the flux density be recovered. If the
source is not well--resolved by the telescope beam, then the total flux
per beam must be evaluated and compared with the mass sensitivity of
the observation over the typical total source linewidth.  These
considerations are crucial in assessing the detectability of a
population of CHVCs when viewed at large distances.

\section{Source symmetries and asymmetries}
\label{sec:sym}

The sensitive, high-resolution cross-cuts described in
\S\ref{sec:results} provide the opportunity for studying the edge
profiles of each source in our sample down to very low column densities
as well as their degree of reflection-symmetry. Since only a single
positional angle (along a line of constant declination) has been
observed to this depth in each source we are very sensitive to
the particular sub-structures we happen to encounter. If instead we
were able to employ azimuthal averaging from a fully-sampled large-area
map, we would expect such sub-structures to average away to a large
extent. Nonetheless, some trends are worthy of discussion on the basis
of this limited source sampling.

Firstly, the high column density regions (\NH$>10^{18.5}$cm$^{-2}$) of
each source, which we term cores, show a high degree of
structure. Typically, several of such cores are found in each source,
but even if there is only one prominent core component it is not
necessarily accurately centered within the diffuse low column density
halo.  Prominent cores which we sample that are substantially
off-center are seen in CHVC\,100$-$49$-$383, CHVC\,148$-$32$-$144 and
CHVC\,204$+$30$+$075. In these cases, the two degree scan length of our
deep cross-cut does not extend to both edges of the source, making it
impossible to comment on the degree of symmetry seen in the underlying
halo component. Even in less extreme cases, our spatial coverage is
sometimes insufficient to extend beyond the range of detectable
emission on at least one side of the source, so that comparisons can
only be made over a limited range of column densities. 

The comparison of the ``East'' and ``West'' \NH\ profiles in
Figs.\ref{fig:nh1}--\ref{fig:nh2} is shown for the range of column
densities that is above the noise floor ($>2\sigma$) in each case. The
origin of the ``radius'' axis in these figures was arbitrarily chosen
to correspond to the location of peak column density along the single
cross-cut. Positive or negative shifts in position are therefore
allowed, and source symmetry should be judged on the basis of agreement
in the local slope of the two profiles rather than on their ``radial''
position. With these caveats in mind, it becomes clear that substantial
asymmetries appear to be confined to the regions of moderately high
column density (\NH$>10^{18.5}$cm$^{-2}$), while below this column
density a high degree of reflection-symmetry is present in all
cases. Even those sources that have extreme asymmetries at
\NH~$>~10^{19}$cm$^{-2}$, like CHVC\,186$-$31$-$206 and
CHVC\,198$-$12$-$103, have effectively identical edge profiles below
\NH~$>~10^{18.2}$cm$^{-2}$.

The large disparity in source symmetry at high and low \NH\ seen in
some sources, particularly in CHVC\,198$-$12$-$103, has important
implications for the physical conditions in and around these
sources. While substantial asymmetry of the high \NH\ regions might be
interpreted as implying an externally induced ram pressure origin
(eg. Br\"uns et al. (\cite{brun00}), this seems to be ruled out by the
high degree of symmetry seen in the low \NH\ envelopes. If the core
asymmetries were due to such an external influence, then the
asymmetries should be even more severe in the diffuse halos, which
clearly is not the case.

Significant asymmetries at high column density levels are also commonly
observed in external galaxies. Richter \& Sancisi (\cite{rich94})
concluded that at least 50\% of all spiral galaxies have strong or mild
asymmetries based on analysis of their integrated \hi\ line
profiles. Swaters (\cite{swat99}) extended this analysis with the
assessment of asymmetry in a sample of 73 late-type dwarf galaxies. He
finds strong asymmetires in 34\% of his sample and weak asymmetries in
a further 16\%. The incidence of significant asymmetry at high \NH\ in
our sample of CHVCs is about 6 out of 10 based on our limited spatial
sampling. This is quite comparable to the rate of incidence seen in
nearby galaxies.
 
\section{3--D morphology and distance}
\label{sec:morph}

The reflection-symmetry of CHVC edge profiles at low column densities
together with the roughly circular appearance of each source at low
angular resolution ($\sim$30~arcmin) imply that the diffuse halo
component may have a substantial degree of spherical symmetry. These
properties suggest a method to constrain the three--dimensional
morphology and possible distances of these objects. If we consider an
intrinsically exponential distribution of atomic volume density in the
diffuse WNM halos of the CHVCs with spherical symmetry of the form
\begin{equation}
n_{\rm H}(r) = n_{\rm o}e^{-r/h_B}
\label{eqn:no}
\end{equation}
in terms of the radial distance, $r$, and exponential scale length,
$h_B$, it is possible to calculate the corresponding projected distribution of
\hi column density,
\begin{equation}
N_{\rm HI}(r) = 2 h_B n_{\rm o} \biggl[ {r \over h}K_1 \biggl( { r \over
h_B}\biggr)
\biggr],
\label{eqn:No}
\end{equation}
where $K_1$ is the modified Bessel function of order 1. This result
follows from the related calculation of the edge--on appearance of an
exponential stellar disk by Van der Kruit \& Searle (\cite{vand81}).

The projected distribution of \NH\ given by eqn.~\ref{eqn:No} is
approximately exponential beyond a few scale--lengths, but flattens
significantly toward small radii. The correspondence between these
Bessel function scaleheights, $h_B$, and the most similar 1-D
exponential scaleheights, $h_e$, (over the interval $3h_B<r<6h_B$) is
about $h_e$~=~1.1~$h_B$.  \NH\ profiles of the form given by
eqn.\ref{eqn:No} have been overlaid on the data shown in
Figs.~\ref{fig:nh1} and \ref{fig:nh2}. Profiles of this type provide a
reasonably good description of the observed profiles and allow accurate
assessment of the total atomic column density, \NH(0), and intrinsic
scale--length, $h_B$, of the WNM halos under the assumption of crude
spherical symmetry. Even in those cases where there is some asymmetry
between the Eastern and Western halves of the profile at high \NH, the
peak \NH\ value of the halo and the scale--length at large radii are
well--defined.  Due to the large offset of the bright core location
from the centroid of the underlying halo in CHVC\,100$-$49$-$383,
CHVC\,148$-$32$-$144, and CHVC\,204+30+075 only one of the edge
profiles was observed in our cross-cut. However, even for these
sources, the well-sampled edge of the source shows good correspondence
with profiles of this form.  For the sources CHVC\,092$-$39$-$367,
CHVC\,202$+$30$+$057, and CHVC\,204$+$30$+$075, the calculated \NH\
profiles were given a linear offset from the origin in radius, since
the transition from cool cores to warm halos was significantly offset
from the direction that displayed the peak column density. The
displayed values of the total atomic column density, \NH(0), and
scale-length, $h_B$, are listed in Table~\ref{tab:results} for the
seven sources in which both Eastern and Western edge profiles were
sampled.  Significantly worse correspondence of the profiles with the
data is found for variations of 0.1 dex in log(\NH) and 50~arcsec in
$h$.  We find mean values of \NH(0)$ = 4.1\pm 3.2 \times
10^{19}$\,cm$^{-2}$ and $h = 420 \pm 90$ arcsec averaged over the 7
tabulated objects.

BB00 considered the physical conditions necessary for the shielding and
condensation of CNM cores within WNM halos. While thermal pressures,
$P/{\rm k}$, of $\sim$2000~cm$^{-3}$\,K are found in the local
mid--plane of the Galaxy, these are expected to decline dramatically
with height (and Galactocentric radius), falling to values below about
100~cm$^{-3}$\,K beyond about 20~kpc (Wolfire et al. \cite{wolf95b}).
The calculations presented in Wolfire et al. suggest that for a wide
range of physical conditions (metal abundance, radiation field and dust
content) the transition from a two phase ISM to a WNM should occur near
a thermal pressure of $P/{\rm k} \sim$~100~cm$^{-3}$\,K. Only in the
case of effectively primordial metal abundance is a large departure
expected from this nominal transition pressure, in the sense of a much
higher required thermal pressure. From the ubiquitous detection of
metal line systems in quasar absorption line studies as well as a
metallicity estimate for CHVC125+41$-$207 (BB00) it seems likely
that a metal abundance of about 0.1 solar is appropriate for the CHVCs.
Assuming a nominal thermal pressure of the core/halo interface in CHVCs
allows calculation of the central volume density, $n_o$, since the
kinetic temperature is known from the observed linewidths to be $T_{\rm
k}$~=~10$^4$\,K. The distance of each object with a measured edge
profile can then be estimated by assuming an equal extent in the plane
of the sky and along the line-of-sight from,
\begin{equation}
D = {N_{\rm HI}(0) \over 2 h n_o} = {N_{\rm HI}(0) {\rm k} T \over 2 h P}, 
\label{eqn:D}
\end{equation}
or
\begin{equation}
D =  335 \biggl({N_{\rm HI}(0) \over 10^{19}\,{\rm cm}^{-2}}\biggr)
\biggl({P/{\rm k} \over 100\,{\rm cm}^{-3}\,{\rm K}}\biggr)^{-1}  \biggl({h
\over 100''}\biggr)^{-1}  {\rm kpc.}
\label{eqn:Dis}
\end{equation}
These estimated distances are listed in Table~\ref{tab:results} and
vary between 150 and 850~kpc, with a mean of 400$\pm$280~kpc. 

A similar method of distance estimation for HVCs was suggested by
Ferrara \& Field (\cite{ferr94}), although in their treatment a uniform
volume density of the gas was assumed, which complicates the
identification of the angular size that should be related to the
line--of--sight depth.  The distances to several of the diffuse HVC
complexes have been found, using more direct methods (e.g. van Woerden
et al. \cite{woer99}), to lie in the range of several to several tens
of kpc.  But the distinctions between the diffuse HVC complexes and the
population of compact, isolated CHVCs are sufficiently robust, and the
distance parameter so crucial, that investigations of alternative
measures of distance are required.  We note that the WSRT
interferometric imaging carried out by Wakker \& Schwarz
(\cite{wakk91c}) towards two CHVCs and along representative
lines--of--sight towards four HVC complexes revealed that the two
object classes have very different properties. A much larger fraction
of the object flux was recovered in the WSRT imaging of the CHVCs and
the detected cool cores are much more subject to beam dilution effects
when observed with a coarser beam. Both of these properties led Wakker
\& Schwarz to conclude that the CHVCs were significantly more distant
than the HVC complexes.

It is important to stress that we have considered only the distribution
of the observed atomic gas density in our analysis. It is likely
that the edges of the atomic distribution are accompanied by an
increasing proportion of ionized gas due to photoionization by external
radiation. For this reason, the total gas density (the sum of atomic
and ionized components) will probably be significantly shallower than
the distributions shown here. However, the method of distance
estimation discussed above is not effected in a systematic way by
considering only the atomic rather than the total gas distribution. We
only assume an approximate spherical symmetry of the atomic component,
which is still expected to be the case.

\begin{figure*}[h]
\resizebox{18cm}{!}{\includegraphics{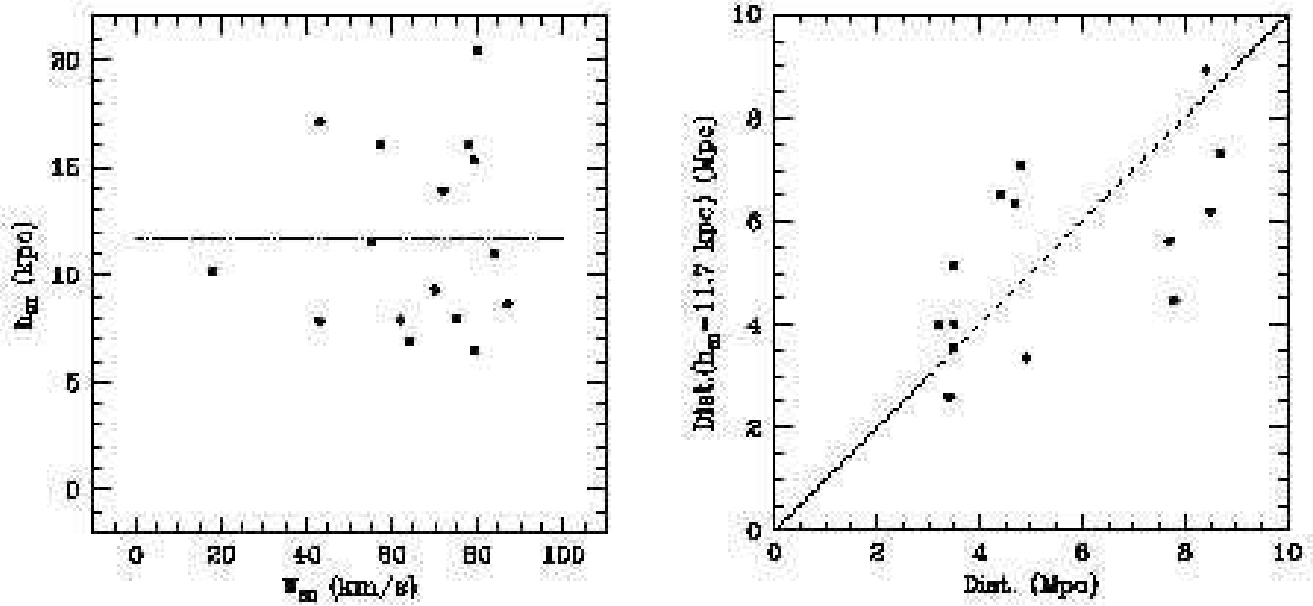}}
\caption{ Outer--disk exponential scale--length of \hi as a function of
profile half--width in a sample of 15 nearby late--type dwarf galaxies
taken from Swaters (\cite{swat99}) is indicated in the left-hand
panel.  The dotted line is the mean value, $h_e$~=~11.7$\pm$4.4~kpc. In
the right--hand panel, the distance of the sample galaxies,
derived assuming a constant intrinsic scale--length, $h_e$~=~11.7~kpc, is
plotted against the optically measured distance. Equal distances would
follow the dotted line. The actual distances are returned with less than
a factor of two scatter by assuming $h_e$~=~11.7~kpc. }
\label{fig:swh}
\end{figure*}

We can calibrate the edge-profile method of distance determination by
direct comparison with the observed edge--profiles of low mass
galaxies. A comprehensive study of 73 late--type dwarf galaxies based
on a uniform database has recently been presented by Swaters
(\cite{swat99}). Optical and \hi\ properties, including the exponential
scale-lengths of \hi\ in the outer disks are derived for this entire
sample. We have taken all of the well--resolved, high signal--to--noise
data for the lowest mass systems studied by Swaters by selecting the
galaxies nearer than 10~Mpc, with line profile half--widths ($W_{50}$)
less than 100~\kms, and average \hi\ surface density within 3.2 optical
disk scale--lengths ($<\Sigma_{\rm HI}>_{3.2h}$) greater than 2.5
M$_\odot\,$pc$^{-2}$. The exponential scale--length fit to the outer disk
\hi\ for the 15 selected galaxies is plotted against the line profile
half-width in the left--hand panel of Fig.\,\ref{fig:swh}. There is only
modest variation in the outer disk scale--lengths and no apparent
variation with profile line--width. The mean outer disk scale-length is
$h_e$~=~11.7$\pm$4.4~kpc. If we adopt this mean outer disk scale-length
as a constant for all low mass galaxies and calculate the corresponding
distances implied by $h_e$ in angular units, we recover the accepted
galaxy distances to better than a factor of two, as shown in the
right-hand panel of Fig.\ref{fig:swh}. The corresponding value of $h_B$
for a Bessel function scale--length is $h_B$~=~10.6$\pm$4.0~kpc.
Applying this same yard--stick to our measurements of halo scale--length
in the CHVCs yields the distance estimates in the last column of
Table~\ref{tab:results}. The derived distances vary between 320 and
730~kpc, with a mean of 500$\pm$130 kpc, and show agreement to within
about a factor of two with our estimates from eqn.\,\ref{eqn:Dis}.

\section{Discussion and conclusions}
\label{sec:disc}

The Arecibo imaging reported here has demonstrated that a nested
core/halo geometry is characteristic of the targeted CHVCs, and allows
the direct resolved detection of the halo component for the first time
in these compact objects.  The compact cores identified in our earlier
WSRT interferometric data were only marginally resolved at angular
resolutions of $2'$ or less; such cores would remain severely unresolved
in the Arecibo beam. On the other hand, the signal from diffuse halo
material is effectively undetected in the synthesis data and can only be
studied by deep total--power imaging.

The measured FWHM linewidths in the halos are typically in the range 20
to 25 \kms. The various circumstances, in particular blending, which
would render this measure an upper limit to the thermal broadening are
shown to be not especially important by the accuracy of the
single--Gaussian spectral fits, particularly in the wings of the
spectra.  The residuals of the single--Gaussian fits are in some cases
in excess of the {\sc rms} noise of some 10 mK at a velocity resolution of
1~\kms.  However, the dimples evident in the profiles imply
unresolved structure in the ensemble of cores. The general shape of the
profile, and particularly the form of the spectral wings, is consistent
with thermal emission from \hi at a kinetic temperature near $10^4$ K.
Thus these observations provide evidence that the 
halos of the CHVCs observed are comprised of gas in the WNM phase,
hypothesized as providing the neutral shielding column required if the CNM
cores are to remain stable in the presence of an ionizing radiation field.

A kinematic gradient aligned with the elongation of the object was
observed in several CHVC cores, most notably in CHVC\,158$-$39$-$285
and CHVC\,186$-$31$-$206.  Since the long--integration Arecibo data was
limited to a single East--West cross--cut that only rarely coincides
with the direction of elongation, this conclusion can not be stated
firmly for all of the objects observed. The kinematic gradients are
compatible with the expectation of slow rotation in a flattened disk
system, and are consistent with the modeling which BB00 applied to two
well--resolved cores in CHVC\,204+30+075, showing that these cases can
be described by rotation curves of Navarro et al.  (\cite{nava97}) cold
dark matter halos of mass about $10^8$ M$_\odot$.

The halos and the cores are evidently kinematically decoupled, in the
sense that a core region may show a systematic kinematic gradient when
the halo does not, but remains centered on the systemic velocity of the
CHVC.  CHVC\,158$-$39$-$285 is a particular example of this situation
(although the same effect can also be seen in CHVC\,100$-$48$-$383 and
CHVC\,186$-$31$-$206), which is compatible with rotation but seems
difficult to reconcile with an external cause for the kinematic
distortion, for example by a tidal shear.  An external force would be
expected to produce a more severe distortion of the diffuse envelopes
than on the dense cores. A tidal shear or other externally caused
distortion seems unlikely for several other reasons, including the
apparent isolation of the CHVCs and the evidence for a Local Group
deployment.

Substantial asymmetries characterize the high column density
(\NH$>10^{18.5}$cm$^{-2}$) portions of the \hi distributions in 6 out
of 10 of the CHVCs imaged. Particularly striking examples of this high
\NH\ lopsidedness are provided by CHVC\,100$-$49$-$383,
CHVC\,148$-$32$-$144, and CHVC\,198$-$12$-$103.  This rate of occurence
of asymmetry is comparable to the 50\% found in large samples of nearby
galaxies by Richter \& Sancisi (\cite{rich94}) and Swaters
(\cite{swat99}).  In contrast, the low column density
(\NH$<10^{18.5}$cm$^{-2}$) portions of the edge profiles display a very
high degree of reflection--symmetry in all cases where our spatial
coverage extends to these levels. Even the extreme source,
CHVC\,198$-$12$-$103, displays essentially identical edge profiles
below \NH~$=~10^{18.2}$cm$^{-2}$. The high degree of symmetry found
in the low \NH\ halos relative to the high \NH\ cores argues against
external mechanisms, like ram--pressure confinement, as the cause of the
asymmetries seen in the cores. 

The \NH~profiles of the warm halos plotted in Fig.~\ref{fig:nh1} and
\ref{fig:nh2} can be described well, by the sky--plane projection of a
spherical exponential distribution of \hi volume density. In the most
sensitive cross-cuts the exponential edge profile can be tracked down to
\NH~$=~10^{17.5}$cm$^{-2}$.  Thus, the measured size and total flux of
the CHVCs will be dependent on the resolution and on the sensitivity of
the observation.  This point is of particular concern when considering
searches for CHVC counterparts which might be associated with galaxies
or groups of galaxies beyond the Local Group.

The distinctive halo profiles also allow estimation of the peak halo
column density and exponential scale--length in each object. When
combined with an estimate of the nominal thermal pressure at the
core/halo interface, the distance to each object can be derived by
assuming approximate spherical symmetry. Seven of the ten objects
studied indicate distances which vary between 150 and 850~kpc, with a
mean of 400$\pm$280~kpc. If instead we adopt the mean observed outer
disk exponential scale--length of about 11~kpc found in a sample of
nearby late--type dwarf galaxies, the same set of objects yield
consistent distances that vary between 320 and 730~kpc with a mean of
500$\pm$130~kpc.

\section{Summary}
\label{sec:summ}

The Arecibo imaging of ten CHVCs, combined with the BB00 WSRT imaging, lead
to the following conclusions:

\vspace{.2cm}
\noindent $\bullet$  The compact high--velocity clouds display a 
core/halo morphology.

\noindent $\bullet$  The CNM of the cores of CHVCs are better revealed in 
synthesis data than in filled--aperture data because of their small angular 
scales.

\noindent $\bullet$  The WNM of the diffuse halos of CHVCs are better 
revealed in filled--aperture data than in synthesis data because of their 
diffuse morphology and large angular sizes.

\noindent $\bullet$  The FWHM linewidth of the CHVC halo gas is 
typically 25 \kms, consistent with the expected WNM thermal linewidth of
10$^4$~K gas. 

\noindent $\bullet$  Some of the cores exhibit a kinematic gradient, 
consistent with rotation.

\noindent $\bullet$  The differing kinematics of the cores and the
halos is consistent with the cores being kinematically decoupled from the
halos.

\noindent $\bullet$  Since the halos do not commonly share the kinematic 
gradient of cores, an external origin of the gradient is unlikely.

\noindent $\bullet$  60\% of the CHVCs studied display significant
asymmetries in the \hi distribution at high column density
(\NH$>10^{18.5}$cm$^{-2}$). 

\noindent $\bullet$ There is a high degree of reflection-symmetry in
the low column density (\NH$<10^{18.5}$cm$^{-2}$) edge profiles in all
seven cases where the spatial coverage extends to these levels.

\noindent $\bullet$ The presence of a highly symmetric halo exterior to
an asymmetric core argues against external mechanisms for producing
the core asymmetry.

\noindent $\bullet$ The \hi column--density profiles of the CHVC halos
are consistent with the sky--plane projection of a spherical
exponential distribution of \hi volume density.

\noindent $\bullet$ Assuming a nominal thermal pressure for the CNM/WNM
interface and approximate spherical symmetry of the halos, distance
estimates for individual objects can be made. These lie in the range
150 to 850~kpc.

\noindent $\bullet$ Assuming a ``standard'' outer-disk scale-height for
low mass self-gravitating systems (as observed in nearby dwarf
galaxies) allows distance estimates in the range 320 to 730~kpc to be
made.

\begin{acknowledgements} 
  
  We are grateful to P. Perillat for assistance during our Arecibo
  observing session, to T.M. Bania for communications regarding the
  calibration of the data, to C. Salter and E. Howell for information
  regarding the beamwidth and sidelobe characterisitics, and to the
  referee, J. Kerp, for constructive comments.  The Arecibo Observatory
  is part of the National Astronomy and Ionosphere Center, which is
  operated by Cornell University under a cooperative agreement with the
  National Science Foundation.

\end{acknowledgements}


\end{document}